%% file: neurips_data_2021.tex
\documentclass{article}

\usepackage[preprint]{neurips_data_2021}

\usepackage[utf8]{inputenc} 
\usepackage[T1]{fontenc}    
\usepackage[colorlinks=True, 
            linkcolor = blue,
            urlcolor  = blue,
            citecolor = black,]{hyperref}       
\usepackage{url}            
\usepackage{booktabs}       
\usepackage[table,xcdraw,dvipsnames]{xcolor}
\usepackage{amsfonts}       
\usepackage{nicefrac}       
\usepackage{microtype}      
\usepackage{xcolor}         

\usepackage{graphicx}
\usepackage{caption}
\usepackage{subcaption}

\definecolor{codegreen}{rgb}{0,0.6,0}
\definecolor{codegray}{rgb}{0.5,0.5,0.5}
\definecolor{codepurple}{HTML}{C42043}
\definecolor{backcolour}{HTML}{F2F2F2}
\definecolor{bookColor}{cmyk}{0,0,0,0.90}  

\usepackage{listings}
\usepackage{textcomp}
\lstdefinestyle{mystyle}{
    backgroundcolor=\color{backcolour},   
    commentstyle=\color{codegreen},
    keywordstyle=\color{codepurple},
    numberstyle=\numberstyle,
    stringstyle=\color{codepurple},
    basicstyle=\ttfamily\color{bookColor},
    breakatwhitespace=false,
    breaklines=true,
    captionpos=b,
    keepspaces=true,
    numbers=left,
    numbersep=10pt,
    showspaces=false,
    showstringspaces=false,
    showtabs=false,
}
\usepackage{multirow}
\usepackage[first=0,last=9]{lcg}
\definecolor{Gray}{gray}{0.9}
\usepackage{placeins} 
\usepackage{dirtree} 

\usepackage[disable]{todonotes}

\newcommand{\ikn}[1]{\todo[size=\scriptsize, color=orange]{[IK] #1}}

\newcommand{\yln}[1]{\todo[size=\scriptsize, color=blue!20]{[YL] #1}}

\usepackage{quoting,xparse}
\NewDocumentCommand{\bywhom}{m}{
  {\nobreak\hfill\penalty50\hskip1em\null\nobreak
   \hfill\mbox{\normalfont(#1)}%
   \parfillskip=0pt \finalhyphendemerits=0 \par}%
}

\NewDocumentEnvironment{pquotation}{m}
  {\begin{quoting}[
     indentfirst=true,
     leftmargin=\parindent,
     rightmargin=\parindent]\itshape}
  {\bywhom{#1}\end{quoting}}

\usepackage{pifont}
\title{HumBugDB: A Large-scale Acoustic Mosquito Dataset}
\author{%
  Ivan Kiskin\footnotemark[1]  \\
  University of Oxford
  \And
  Marianne Sinka\footnotemark[2] \\
  University of Oxford
  \And
  Adam D. Cobb\footnotemark[6] \\
  SRI International
     \And
      Waqas Rafique\footnotemark[1]\\
  University of Oxford
     \And
  Lawrence Wang\footnotemark[1]\\
  University of Oxford
  \And
         Davide Zilli\footnotemark[5] \\
  Mind Foundry Ltd
      \And
         Benjamin Gutteridge\footnotemark[1]\\
  University of Oxford
      \And
          Rinita Dam\footnotemark[2]\\
  University of Oxford
    \And
    Theodoros Marinos\footnotemark[8]\\
    University of Surrey
    \And
    Yunpeng Li\footnotemark[8]\\
    University of Surrey
      \And
  Dickson Msaky\footnotemark[3] \\
  IHI Tanzania
     \And
      Emmanuel Kaindoa\footnotemark[3] \\
  IHI Tanzania
     \And
  Gerard Killeen\footnotemark[4]\\
  UCC, BEES
     \And
    Eva Herreros-Moya\footnotemark[2]\\
     University of Oxford
 \\
     \And
  Kathy Willis\footnotemark[2]\\
  University of Oxford
     \And
  Stephen J. Roberts\footnotemark[1]\\
  University of Oxford
\AND
\\
\footnotemark[1]\enspace \small \texttt{\{ikiskin, waqas, beng, sjrob\}@robots.ox.ac.uk}, \small \texttt{lawrence.wang@eng.ox.ac.uk} \\ \small \footnotemark[2]\enspace  \texttt{\{marianne.sinka, kathy.willis, rinita.dam, eva.herreros-moya\}@zoo.ox.ac.uk}, \\ \small \footnotemark[6]\enspace\texttt{adam.cobb@sri.com},
\footnotemark[8]\enspace \, \texttt{\{tm00591, yunpeng.li\}@surrey.ac.uk}, \footnotemark[4]\enspace \texttt{gerard.killeen@ucc.ie},\\\small \footnotemark[5]\enspace \texttt{davide.zilli@mindfoundry.ai}, 
\footnotemark[3]\enspace  \texttt{\{dmsaky,ekaindoa\}@ihi.or.tz}.
}

\begin{document}

\include{neurips_data_2021_main}
\include{neurips_data_2021_supplement}



\end{document}

%% file: neurips_data_2021_main.tex
\newcolumntype{g}{>{\columncolor{Gray}}c}
\newcolumntype{G}{>{\columncolor{Gray}}l}
\newcolumntype{R}{>{\columncolor{Gray}}r}

\maketitle
\begin{abstract}
This paper presents the first large-scale multi-species dataset of acoustic recordings of mosquitoes tracked continuously in free flight. We present 20 hours of audio recordings that we have expertly labelled and tagged precisely in time. Significantly, 18 hours of recordings contain annotations from 36 different species. Mosquitoes are well-known carriers of diseases such as malaria, dengue and yellow fever. Collecting this dataset is motivated by the need to assist applications which utilise mosquito acoustics to conduct surveys to help predict outbreaks and inform intervention policy. The task of detecting mosquitoes from the sound of their wingbeats is challenging due to the difficulty in collecting recordings from realistic scenarios. To address this, as part of the HumBug project, we conducted global experiments to record mosquitoes ranging from those bred in culture cages to mosquitoes captured in the wild. Consequently, the audio recordings vary in signal-to-noise ratio and contain a broad range of indoor and outdoor background environments from Tanzania, Thailand, Kenya, the USA and the UK. In this paper we describe in detail how we collected, labelled and curated the data.  The data is provided from a PostgreSQL database, which contains important metadata such as the capture method, age, feeding status and gender of the mosquitoes. Additionally, we provide code to extract features and train Bayesian convolutional neural networks for two key tasks: the identification of mosquitoes from their corresponding background environments, and the classification of detected mosquitoes into species. Our extensive dataset is both challenging to machine learning researchers focusing on acoustic identification, and critical to entomologists, geo-spatial modellers and other domain experts to understand mosquito behaviour, model their distribution, and manage the threat they pose to humans.
\end{abstract}

\section{Introduction}
\label{sec:Introduction}

There are over 100 genera of mosquito in the world containing over 3,500 species and they are found on every continent except Antarctica \citep{harbach2013mosquito}. Only one genus (\textit{Anopheles}) contains species capable of transmitting the parasites responsible for human malaria. \textit{Anopheles} contain over 475 formally recognised species, of which approximately 75 are vectors of human malaria, and around 40 are considered truly dangerous \citep{sinka2012global}. These 40 species are inadvertently responsible for more human deaths than any other creature. In 2019, for example, malaria caused around 229 million cases of disease across more than 100 countries resulting in an estimated 409,000 deaths \citep{world2020world}. It is imperative therefore to accurately locate and identify the few dangerous mosquito species amongst the many benign ones to achieve efficient mosquito control.
Mosquito surveys are used to establish vector species' composition and abundance, human biting rates and thus the potential to transmit a pathogen. Traditional survey methods, such as human landing catches, which collect mosquitoes as they land on the exposed skin of a collector, can be time consuming, expensive, and are limited in the number of sites they can survey. They can also be subject to collector bias, either due to variability in the skill or experience of the collector, or in their inherent attractiveness to local mosquito fauna. These surveys can also expose collectors to disease. Moreover, once the mosquitoes are collected, the specimens still need to undergo post sampling processing for accurate species identification. Consequently, an affordable automated survey method that detects, identifies and counts mosquitoes could generate unprecedented levels of high-quality occurrence and abundance data over spatial and temporal scales currently difficult to achieve. \ikn{New section until end of paragraph} 
We therefore utilise low-cost smartphones, acting as acoustic mosquito sensors, to solve this task. The exponential increase in smartphone ownership is a worldwide phenomenon. Governments and independent companies are continuing to extend connectivity across the African continent \citep{friederici2017impact}. More than half of sub-Saharan Africa is expected to be connected to a mobile service by 2025 \citep{global2020mobile}. With this expanding coverage of mobile phone networks across Africa, there is an emerging opportunity to collect huge datasets, as exemplified by the World Bank's Listening to Africa Initiative \citep{worldbank2017}. Our target application (Section \ref{sec:humbug}) uses a free downloadable app, which means that every smartphone can be a mosquito monitor.

 

\paragraph{Our contribution}  
In order to assist research in methods utilising the acoustic properties of mosquitoes, as part of the HumBug project (described in Section \ref{sec:humbug}) we contribute:

\begin{itemize} 
\item \textbf{Data:} \url{http://doi.org/10.5281/zenodo.4904800}: A vast database of 20 hours of finely labelled mosquito sounds, and 15 hours of associated non-mosquito control data, constructed from carefully defined recording paradigms. Data was collected over the course of five years in a global collaboration with  mosquito entomologists. Recordings were captured from 36 species 
with a mix of low-cost smartphones and professional-grade recording devices, to capture both the most accurate noise-free representation, as well as the sound that is likely to be recorded in areas most in need. \ikn{new sentence}A diverse quantity of wild and lab culture mosquitoes is included in the database to capture the biodiversity of naturally occurring species.
Our data is stored and maintained in a PostgreSQL database, ensuring label correctness, data integrity, and allowing efficient updates and re-release of data.


\item \textbf{Mosquito event detector and species classification baselines:} \url{https://github.com/HumBug-Mosquito/HumBugDB}: Detailed tutorial code for training state-of-the-art  Bayesian neural network models for two key tasks -- Mosquito Event Detection (MED): distinguishing mosquitoes of any species from their background surroundings, such as other insects, speech, urban, and rural noise; Mosquito Species Classification (MSC): species classification of over 1,000 individually captured wild mosquitoes.  In combination, our tasks and models are the first of their kind to use large-scale real-world data for the purpose of automating acoustic mosquito species monitoring. 
\end{itemize}

The rest of the paper is structured as follows. Section \ref{sec:relatedwork} details related datasets and describes how ours contributes to the literature uniquely. Section \ref{sec:humbug} shows the  intended use cases for the data and models released in this paper.
Section \ref{sec:Datacollection} describes in depth the sources and collection methods of data present. The steps taken to benchmark models for MED and MSC are given in Section \ref{sec:benchmark}. We discuss the results that our models achieve, and the open challenges remaining. We conclude in Section \ref{sec:conclusion}. 

\ikn{New overview}Comprehensive instructions for using our baseline models and feature extraction code are provided in Appendix \ref{sec:appendix_code_use}, and additional details on all the metadata in Appendix \ref{sec:appendix_db_metadata}. The datasheet (Appendix \ref{sec:appendix_datasheet_for_dataset}) details the dataset's composition (\ref{sec:appendix_composition}), acquisition process (\ref{sec:appendix_datasheet_collection}), preprocessing (\ref{sec:appendix_datasheet_preprocessing}), past and suggested use cases (\ref{sec:appendix_datasheet_uses}), data bias and mitigation strategies (\ref{sec:appendix_datasheet_databias}), and maintenance policies (\ref{sec:appendix_database_maintenance}).

\section{Related work}
\label{sec:relatedwork}

Mosquitoes have particularly short, truncated wings allowing them to flap their wings faster than any other insect of equivalent size -- up to 1,000 beats per second \citep{simoes2016role,bomphrey2017smart}. This produces their distinctive flight tone and has led many researchers to try and use their sound to attract, trap or kill them
\citep{perevozkin2015species,johnson2016siren,jakhete2017wingbeat,joshi2021review}. Table \ref{tab:DatasetLiterature} provides details of the few \yln{I feel that ``the few'' means that we are listing all related datasets and there is a danger of missing some. Would it be slightly better to change it to ``several related''} datasets released to the public to aid this research.\ikn{Do some extra research to make sure we didn't miss any} We discuss the varying sensor modalities separately, due to their inherent differences in properties.

\begin{table}[ht]
\caption{Publicly available datasets. \textit{`Average mosquito'} is the approximate length of audible mosquito recording per sample. Where not known, \textit{`Mosquito'} is estimated from the average mosquito sample duration multiplied by the number of positive samples.  \textit{`Type'} represents wild captured or lab grown mosquitoes (in order of prevalence).   Crowdsourced recordings or labels are marked with (\textbf{*}).}
\label{tab:DatasetLiterature}
\centering
\small
\begin{tabular}{@{}llllll@{}}
\toprule
\rowcolor{white}
Dataset                & Sensor & \begin{tabular}[l]{@{}l@{}}Mosquito\\ (Background)\end{tabular}  & \begin{tabular}[l]{@{}l@{}}Average \\ mosquito \end{tabular} & Species & Type \\ \midrule
\citet[UCR]{chen2014flying}             & Opto-acoustic &  17\,min  (N/A)         &  $\approx$\,0.02 s       &       6            &     Lab          \\\rowcolor{white}
\citet{PotamitisKaggle}   & Opto-acoustic    &    39\,hr (N/A)  &    $\approx$ 0.5\,s     &          6         &         Lab          \\
\citet{vasconcelos2020annotated} & Acoustic           &      15\,min  (N/A)    &      0.3\,s     &       3            &         Lab   \\\rowcolor{white}
\citet{Mukundarajan2017}  (\textbf{*})         &     Acoustic         &   N/A (N/A)     &    N/A          &          20  &     Lab + wild \\  \citet{kiskin2019data} (\textbf{*})         &     Acoustic         &   2\,hr  (20\,hr)     &   1\,s           &       N/A  &     Lab + wild   \\ \midrule \rowcolor{white}
\textbf{HumBugDB}              & Acoustic           &  20\,hr  (15\,hr)        &   9.7 s      & 36                &       Wild + lab           \\ \bottomrule
\end{tabular}
\end{table}

\paragraph{Opto-acoustic approaches}
\textit{`Wingbeats'} \citep{PotamitisKaggle} and \textit{`UCR Flying Insect Classification'} \citep{chen2014flying} are datasets collected via optical sensors with high signal-to-noise-ratio (SNR). We note this is a different, but complementary, approach. 
Due to the directionality of the recording method, typical sample durations are encountered from ``only a few hundredths of a second'' \citep{chen2014flying} to approximately half a second \citep{PotamitisKaggle}. The approach therefore does not capture the acoustical properties of mosquito sound in free flight which aid mosquito detection in purely acoustic approaches \citep{vasconcelos2020annotated}. 
Furthermore, these datasets survey lab-grown mosquito colonies which do not capture the biodiversity of mosquitoes encountered in the wild \citep{huho2007nature,10.1093/jee/toy024}.
 
\paragraph{Acoustic approaches}
\cite{vasconcelos2020annotated} motivated their release by stating that none of the published datasets include environmental noise, which is essential to fully characterise mosquitoes in real-world scenarios. The dataset consists of 300\,ms snippets, amounting to 15 minutes of recordings. This is an excellent first step. However, for deep learning algorithms the dataset is not readily useable due to its size. Moreover, state-of-the-art models for acoustic classification use training example sizes of at least 0.96 seconds for a variety of audio event detection tasks \citep{hershey2017cnn, pons2017end, Shimada2020_task3_report}.
\ikn{Reference diversity} Our dataset consists of mosquito samples with an average duration of 10 seconds. Additionally, we supply equal quantities of background collected in the same controlled conditions to form a balanced class distribution of mosquito occurrences and a negative control group (see Section \ref{sec:Datacollection}). This is to prevent the recording device or background environment from becoming a confounding factor for the detection of acoustic events \citep{coppock2021covid}. \ikn{Emphasised the need for a negative control group}

\citet{Mukundarajan2017} released an acoustic dataset recorded in free flight with smartphones. However, due to a lack of a rigorous protocol, the quality of the recordings is inconsistent, and there is a lack of metadata recording external factors which influence mosquito sound. There are no labels to timestamp the mosquito events in files where mosquito sound is only sporadic, detracting from the overall utility of the dataset. 

\citet{kiskin2019data} released 22 hours of audio, with crowdsourced labels covering overlapping two-second sections. However, of these, only 2 hours were labelled as containing mosquito sound. In addition, the accuracy of the labels was unknown, and the task of labelling was made difficult as clips were presented in isolation, lacking the relevant background information that specialists utilised for their labels. Curated data of that release is a subset of HumBugDB, in which we improve upon the past release thanks to a joint effort between the zoological and machine learning communities.

Nevertheless, we stress that experimentation which combines information from all of the datasets found in the literature is highly encouraged, and may help find solutions that cover multiple recording modalities, such as both opto-acoustic and acoustic sensors.


\section{Data for mosquito-borne disease prevention}
\label{sec:humbug}

\begin{figure}
    \centering
    \includegraphics[trim= 10 285 0 0, clip, width=0.9\textwidth]{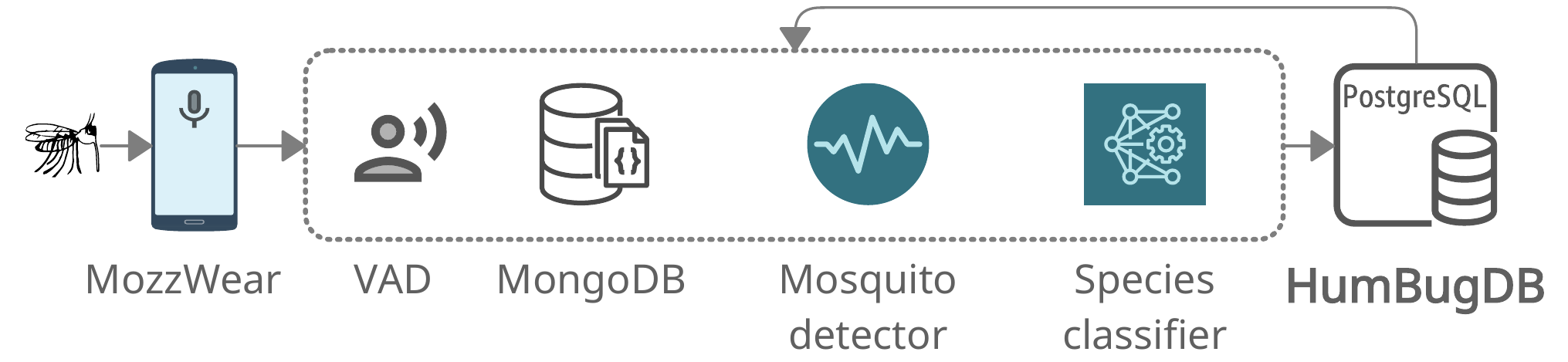}
    \caption{Target workflow. Our mobile phone app, MozzWear, captures audio. The app synchronises to a central server (dashed). Voice activity is removed and data is stored in a MongoDB instance. Audio undergoes mosquito event detection (MED) and subsequent species classification (MSC). Successful detections are used to update HumBugDB. Information feeds back to improve the model.}
    \label{fig:humbugworkflow}
\end{figure}

The HumBug project is a collaboration between the University of Oxford and mosquito entomologists worldwide \citep{humbug}. One of the goals of the project is to develop a mosquito acoustic sensor that can be deployed into the homes of people in malaria-endemic areas to help monitor and identify the mosquito species, allowing targeted and effective vector control.
In the following paragraphs we describe the system of Figure \ref{fig:humbugworkflow} to be deployed for this purpose, the role of each component, and the two key tasks (MED, MSC) our models are able to address thanks to the data of HumBugDB.

\paragraph{Capturing mosquito with smartphones} 
We developed a power-efficient app to record mosquito flight tone using the in-built microphone on a smartphone (MozzWear \citep{mozzwear}). We used 16-bit mono PCM wave audio sampled at 8,000\,Hz, based on prior acoustic low-cost smartphone recording solutions for mosquitoes \citep{li2017mosquito,kiskin2018bioacoustic}.
To ensure mosquitoes fly close enough to a smartphone, we have developed an adapted bednet (the \textit{HumBug Net}) that exploits the inherent behaviour of host-seeking mosquitoes (Figure \ref{fig:mosquito_map}, for details refer to \ikn{Made link clearer @reviewer} \citet[Sec.\,2.1.2]{2021MEE}). The combination of the bednets and smartphones constitutes the intended use case, for which we construct MED: Test A (see Table \ref{tab:Summary}).

\paragraph{MongoDB}
Following app recording, audio is synchronised by the app to a central file server for the storage of sound recordings, and a MongoDB \citep{MongoDB} instance for the storage of metadata. The server possesses a frontend dashboard where recordings and predictions fed back from the model can be accessed. The unstructured nature of the NoSQL engine allows for additional flexibility in storing metadata, especially when new information becomes available.

\paragraph{Mosquito Event Detection (MED)}
A Bayesian convolutional neural network (BCNN), which provides predictions with uncertainty metrics \citep{2021pkddBNN} is used to detect mosquito events. Positive predictions are then filtered by the probability, mutual information and predictive entropy \citep{houlsby2011bayesian}, screened, and stored in a curated database. This drastically reduces the time spent labelling by domain experts -- for our bednet data recorded in Tanzania, we estimate 1 to 2\,\% of 2,000 hours of recorded data contained mosquito events. Finding these events without assistance from the model was infeasible due to the vast quantity of data. Section \ref{sec:Task1_MED} defines two test sets to further motivate model development for this task.

\paragraph{Mosquito Species Classification (MSC)}\ikn{New}
A second BCNN is trained specifically for species classification. Once mosquito events have been identified, a probability distribution over species is produced. The report is made available through an HTML dashboard and can be streamed to the app to provide feedback to users.  Section \ref{sec:Task2_Species} details the MSC task.

\paragraph{PostgreSQL database}
Due to the complex requirements of variables and data storage, we designed a relational database \citep{PSQL} which ensures a standardisation in the labelling and metadata process. 
This mitigates a major cause of data quality issues and time costs in field studies. Data has been obtained from controlled studies in focused experiments, with the aid of MED models where applicable. We discuss the sources of the data present in Section \ref{sec:Datacollection}. Recordings are stored in wave format at their respective sample rates, and all the metadata in \texttt{csv} format (Appendix \ref{sec:appendix_db_metadata}). For our maintenance policy, details of ethics agreements, and detailed documentation, refer to the datasheet (Appendix \ref{sec:appendix_datasheet_for_dataset}).

\paragraph{Privacy}
As a subset of data from the database may contain human speech, and other types of personal data
, we include in this paper only audio which has been assigned an explicit label of \textit{`mosquito', `audio', `background'}, or otherwise full consent from members was obtained (for example where entomology experts state a recording ID). To ensure no speech that has not had explicit consent for is included in future releases, we perform voice activity detection (VAD) and removal using Google's WebRTC project, which is open-source and lightweight \citep{ali2018real,2021webrtc}.
\citet{sahoo2020voice} tested the WebRTC VAD method over 396 hours of data, across multiple recording types.
The approach was between 77\,\% and 99.8\,\% accurate. A list of approved ethical review processes is given in Appendix \ref{sec:appendix_datasheet_collection}.


\section{The HumBugDB dataset}
\label{sec:Datacollection}

\begin{figure}
    \centering
    \includegraphics[trim = 85 31 50 133, clip, width=0.95\textwidth]{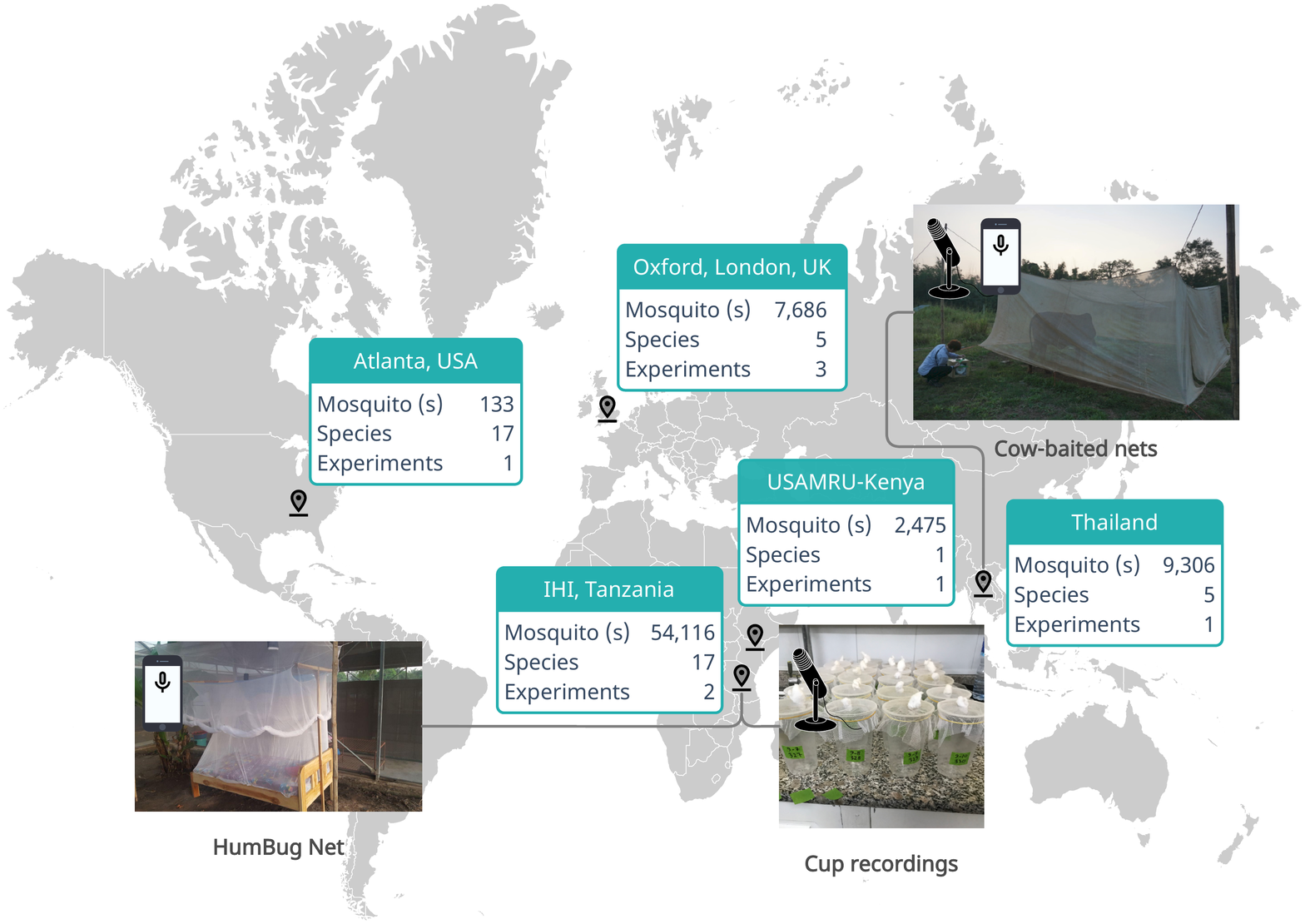}
    \caption{Map of aggregated data acquisition sites. HumBug Net: \citet[Sec.\,2.1.2]{2021MEE}.}
    \label{fig:mosquito_map}
\end{figure}

Our large-scale multi-species dataset contains recordings of mosquitoes collected from multiple locations globally, as well as via different collection methods. Figure \ref{fig:mosquito_map} shows the different locations, with the availability of labelled mosquito sound (in seconds) and number of species, and the number of experiments conducted at each location. In total, we present 71,286 seconds (20 hours) of labelled mosquito data with 53,227 seconds (15 hours) of corresponding background noise to aid with the scientific assessment process, recorded at the sites of 8 experiments.    Of these, 64,843 seconds contain species metadata, consisting of 36 species (or species complexes) with the distributions illustrated in Appendix \ref{sec:appendix_db_metadata}, Figure \ref{fig:species_dist} and Table \ref{tab:appendix_species}.\ikn{Added new Table for species} Table \ref{tab:Summary} gives a more detailed summary of the nature of mosquitoes that were captured, and Appendix \ref{sec:appendix_db_metadata} gives a complete explanation of every field in the metadata. We also demonstrate example spectrograms for a variety of mosquito species in Figure \ref{fig:spec_multispecies}, Appendix \ref{sec:appendix_testPerformance}, and supply a tool to play back and visualise audio clips\footnote{\url{https://github.com/HumBug-Mosquito/HumBugDB/blob/master/notebooks/spec_audio_multispecies.ipynb}} (see Figure \ref{fig:datavis_multispecies}, Appendix \ref{sec:appendix_testPerformance}).

In the following section we break down the data sources according to the nature of mosquitoes -- bred within laboratory culture (Section \ref{sec:lab_culture}) or wild (Section \ref{sec:wild}). We discuss the recording device and the environment the free-flying mosquitoes were recorded in: culture cages, cups or in HumBug Nets.
We also state the methods of capture, where applicable,
documented in more detail in Appendix \ref{sec:appendix_db_metadata}.

\begin{table}
\centering
\footnotesize
\caption{Key audio metadata and division into train/test for the tasks of MED: Mosquito Event Detection, and MSC: Mosquito Species Classification.  \textit{`Wild'} mosquitoes  captured and placed into paper \textit{`cups'} or attracted by bait surrounded by \textit{`bednets'}. \textit{`Culture'} mosquitoes bred specifically for research. Total length (in seconds) of mosquito recordings per group given, with the availability of species meta-information in parentheses. Total length of corresponding non-mosquito recordings, with matching environments, given as \textit{`Negative'}. Full metadata documented in Appendix \ref{sec:appendix_db_metadata}.}
\label{tab:Summary}
\begin{tabular}{@{}GGGGGGG@{}}
\toprule
\rowcolor{white} 
\begin{tabular}[l]{@{}l@{}}Tasks:\\ Train/Test\end{tabular} &\begin{tabular}[l]{@{}l@{}}Mosquito\\ origin\end{tabular}                    &  \begin{tabular}[l]{@{}l@{}}Site\\ Country\end{tabular}                                                        & \begin{tabular}[l]{@{}l@{}}Method\\ (year)\end{tabular} & \begin{tabular}[l]{@{}l@{}}Device\\ (sample rate)\end{tabular} & \begin{tabular}[c]{@{}c@{}}Mosquito (s) \\ (with species)\end{tabular} & \begin{tabular}[l]{@{}l@{}}Negative\\ (s)\end{tabular} \\ 
\midrule
 \begin{tabular}[l]{@{}l@{}}MSC: Train/Test \\ MED: Train\end{tabular}  &Wild                     & \begin{tabular}[l]{@{}l@{}}IHI \\ Tanzania\end{tabular}   & \begin{tabular}[l]{@{}l@{}}Cup \\ (2020)\end{tabular}         & \begin{tabular}[l]{@{}l@{}}Telinga\\ 44.1\,kHz\end{tabular}         & \begin{tabular}[l]{@{}l@{}}45,998\\ 45,998\end{tabular}                      &    5,600      \\  \midrule   

\rowcolor{white}
MED: Train  &Wild    & \begin{tabular}[l]{@{}l@{}}Kasetsart  \\ Thailand\end{tabular}  & \begin{tabular}[l]{@{}l@{}}Cup \\ (2018)\end{tabular}                                                & \begin{tabular}[l]{@{}l@{}}Telinga\\ 44.1\,kHz\end{tabular}           & \begin{tabular}[l]{@{}l@{}}9,306\\ 2,869\end{tabular}          & 7,896     \\  
 
MED: Train &Culture &   \begin{tabular}[l]{@{}l@{}}OxZoology\\ UK\end{tabular}       & \begin{tabular}[l]{@{}l@{}}Cup \\ (2017)\end{tabular}          &           \begin{tabular}[l]{@{}l@{}}Telinga\\ 44.1\,kHz\end{tabular}                                                     & \begin{tabular}[l]{@{}l@{}}6,573\\ 6,573\end{tabular}                   & 1,817     \\  

\rowcolor{white}

MED: Train &Culture & \begin{tabular}[l]{@{}l@{}}LSTMH \\  (UK)\end{tabular}         & \begin{tabular}[l]{@{}l@{}}Cup \\ (2018)\end{tabular}         &                              \begin{tabular}[l]{@{}l@{}}Telinga\\ 44.1\,kHz\end{tabular}                                   & \begin{tabular}[l]{@{}l@{}}376\\ 376\end{tabular}                    & 147      \\ 
              MED: Train&Culture       & \begin{tabular}[l]{@{}l@{}}CDC \\ USA\end{tabular}           & \begin{tabular}[l]{@{}l@{}}Cage \\ (2016)\end{tabular}        & \begin{tabular}[l]{@{}l@{}}Phone\\ 8\,kHz\end{tabular}                                                               & \begin{tabular}[l]{@{}l@{}}133\\ 127\end{tabular}         & 1,121     \\ 
                               \rowcolor{white}
              MED: Train &Culture        & \begin{tabular}[l]{@{}l@{}}USAMRU\\  Kenya\end{tabular}            & \begin{tabular}[l]{@{}l@{}}Cage \\ (2016)\end{tabular}     &                   \begin{tabular}[l]{@{}l@{}}Phone\\ 8\,kHz\end{tabular}                                               & \begin{tabular}[l]{@{}l@{}}2,475\\ 2,475\end{tabular}                      &     31,930     \\ \midrule
                        
        MED: Test A  & Culture &  \begin{tabular}[l]{@{}l@{}}IHI \\ Tanzania\end{tabular}   & \begin{tabular}[l]{@{}l@{}}Bednet \\ (2020)\end{tabular}      & \begin{tabular}[l]{@{}l@{}}Phone\\ 8\,kHz\end{tabular}           & \begin{tabular}[l]{@{}l@{}}4,118\\ 4,118\end{tabular}                   & 3,979   \\ 
                  \rowcolor{white}
         MED: Test B  & Culture                  & \begin{tabular}[l]{@{}l@{}} OxZoology\\ UK\end{tabular}     & \begin{tabular}[l]{@{}l@{}}Cage \\ (2016)\end{tabular}      & \begin{tabular}[l]{@{}l@{}}Phone\\ 8\,kHz\end{tabular}                                                              & \begin{tabular}[l]{@{}l@{}}737 \\ 737\end{tabular}                   &  2,307    \\ \midrule   \rowcolor{white}
                            
\multicolumn{5}{g}{\textbf{Total}}                                                      & \begin{tabular}[l]{@{}l@{}}\textbf{71,286}\\ \textbf{64,843}\end{tabular}                      &    \textbf{53,227}      \\ \bottomrule
\end{tabular}
\end{table}

\subsection{Data collection}
\subsubsection{Laboratory culture mosquitoes}
\label{sec:lab_culture}
Many institutes that conduct research into mosquito-borne diseases hold laboratory cultures of common vector species. These include primary malaria vectors (e.g. \textit{An.} \textit{arabiensis}), primary vectors of the dengue virus (\textit{Aedes albopictus}), yellow fever virus (\textit{Aedes aegypti}) and the West Nile  virus (\textit{Culex quinquefasciatus}).
The controlled conditions of laboratory cultures produce uniformly sized fully-developed adult mosquitoes which are used for a variety of purposes, including trialling new insecticides or examining the genome of these insects.

\paragraph{UK, Kenya, USA}
Mosquitoes were recorded by placing a recording device into the culture cages where one or multiple mosquitoes were flying, or by placing individual mosquitoes into large cups and holding these close to the recording devices (denoted by \texttt{device\_type}).  Recordings were captured at the London School of Tropical Medicine and Hygiene (LSTMH), the United States Army Medical Research Unit-Kenya (USAMRU-K), the Center for Diseases Control and Prevention (CDC), Atlanta, as well as with mosquitoes raised from eggs at the Department of Zoology, University of Oxford. 
We reserve one set of these recordings taken in culture cages by Zoology, Oxford, as MED: Test B (Table \ref{tab:Summary}). Past models were able to achieve excellent mosquito detection performance when trained on recordings held out from the same experiment \citep{kiskin2018bioacoustic,kiskin2017mosquito}. In this paper we treat this experiment as disparate from the remaining data, increasing the difficulty of the detection task. 

\paragraph{Tanzania}
To achieve targeted vector control through the deployment in people's homes, we need to be able to passively capture the mosquito's flight tone. Therefore, in our database we include mosquitoes passively recorded in the Ifakara Health Institute's (IHI) semi-field facility, that most closely resembles the intended use of the HumBug system. It is for this reason that a labelled subset (by an expert zoologist with the help of a BCNN) of this data forms MED: Test A (Table \ref{tab:Summary}). The facility houses six chambers containing purpose-built experimental huts, built using traditional methods and mimicking local housing constructions, with grass roofs, open eaves and brick walls. Four different configurations of the HumBug Net \citep{2021MEE}, each with a volunteer sleeping under the net, were set up in four chambers. Budget smartphones were placed in each of the four corners of the HumBug Net (Figure \ref{fig:mosquito_map}). Each night of the study, 200 laboratory cultured \textit{An.} \textit{arabiensis} were released into each of the four huts and the MozzWear app began recording.


\subsubsection{Wild captured mosquitoes}
\label{sec:wild} \ikn{New focus on species classification and past use cases}
Wild mosquitoes naturally exhibit far greater variability and are thus crucial to sample for real-world detection capability assessment. To study how this affects our ability to distinguish different species, we conducted experiments in Thailand and Tanzania.  Recordings made in Thailand were used to demonstrate that flight tone has the potential to distinguish different species \citep{lidcasefast}. In Section \ref{sec:Task2_Species}, we consider an extension with a greater number of species and more rigorous experimental design with data recorded in Tanzania, forming the MSC dataset of Table \ref{tab:Summary}. 

\paragraph{Thailand}
Across the malaria endemic world, Asia has more \textit{dominant} vector species (mosquitoes whose abundance or propensity to bite humans makes them particularly efficient vectors of disease) than anywhere else. 
Mosquitoes were sampled using ABNs (animal-baited nets in Figure \ref{fig:mosquito_map}), human-baited nets (HBNs) and larval collections (LC) over a period of two months during peak mosquito season (May to October 2018). Sampling was conducted in Pu Teuy Village 
at a vector monitoring station owned by the Kasetsart University, Bangkok. The mosquito fauna at this site include a number of dominant vector species, including \textit{An. dirus} and \textit{An. minimus} alongside their siblings \textit{An. baimaii} and \textit{An. harrisoni} respectively (Appendix \ref{sec:appendix_db_metadata}, Figure \ref{fig:species_dist} and Table \ref{tab:appendix_species} show the exact species distribution).
Mosquitoes were collected at night, carefully placed into large sample cups and recorded the following day using a high-spec Telinga EM23 field microphone and a budget smartphone (see Appendix \ref{sec:appendix_datasheet_collection} for device details).

\paragraph{Tanzania}
\ikn{opportunity to shorten}While Asia has the most diverse vectors, sub-Saharan Africa has the most dangerous mosquito species (\textit{An.} \textit{gambiae}), responsible for
the highest transmission of human malaria in the world, and the highest number of deaths \citep{world2020world}.
In collaboration with the IHI, HBNs, larval collections and CDC-LTs (metadata \texttt{method}, Appendix \ref{sec:appendix_db_metadata}) were used to sample wild mosquitoes in the Kilombero Valley, Tanzania, and record them in sample cups in the laboratory. \textit{An. gambiae} and \textit{An. funestus} (another highly dangerous mosquito found across sub-Saharan Africa), are also siblings within their respective species complexes. Thus, standard polymerase chain reaction (PCR) identification techniques \citep{scott1993identification} were used  to fully identify mosquitoes from these groups.\footnote{The database gives the PCR identification within the \texttt{species} column, or the genus/complex if not available.}
For all the cup recordings in Thailand and Tanzania, environmental conditions (temperature, humidity) were monitored throughout the recording process.\ikn{Need to add these to camera ready. I will. Opportunity to talk about data bias?}
The Tanzanian sampling has collected 17 different species
(Figure \ref{fig:species_dist}, Table \ref{tab:appendix_species} show a full breakdown). Example spectrograms are shown for the eight most populated species in Appendix \ref{sec:appendix_testPerformance} Figure \ref{fig:spec_multispecies}.

\section{Benchmark}
\label{sec:benchmark}\ikn{New link of utility}
\ikn{Can we introduce BNNs in one sentence here?}
To showcase the utility of the data, we supply baseline models for MED in Section \ref{sec:Task1_MED}, and MSC in Section \ref{sec:Task2_Species}. For both tasks, we discuss possible data biases arising from species imbalance, mosquito types, and multiple recording devices, and suggest mitigation strategies in Appendix \ref{sec:appendix_datasheet_databias}. Detailed instructions for code use are  given in Appendix \ref{sec:appendix_code_use}.
Further use cases are discussed in Appendix \ref{sec:appendix_datasheet_uses}.

\paragraph{Models} 
BNNs provide estimates of uncertainty, alongside strong supervised classification performance, which is desirable for real-world use cases such as ours. BNNs are also naturally suited to Bayesian decision theory, which benefits decision-making applications with different costs on error types (e.g. \textit{Anopholes} species are more critical to classify correctly) \citep{vadera2021post,cobb2018loss}. We thus supply three benchmark BNN model classes for this dataset, noting that their equivalent deterministic counterparts achieved either equal or marginally worse classification performance. Details of the training hardware, hyperparameters, and modifications to the models are given in Appendix \ref{sec:appendix_baseline_models}. 
\begin{enumerate}
 \item \textbf{MozzBNNv2}: A CNN with four convolutional, two max-pooling, and one fully connected layer augmented with dropout layers (shown in Appendix \ref{sec:appendix_baseline_models}, Figure \ref{fig:BCNN}). Its structure is based on \href{https://github.com/HumBug-Mosquito/MozzBNN}{prior models} that have been successful in assisting domain experts in curating parts of this dataset with uncertainty metrics \citep{2021pkddBNN}.
\item \textbf{ResNet BNN}: ResNet has achieved state-of-the-art performance in audio tasks \citep{palanisamy2020rethinking} motivating its use as a baseline model in this paper. We augment the model with dropout layers in the building blocks to approximate a BNN. We opt to use the pre-trained model for a warm start to the weight approximations.
 \item \textbf{VGGish BNN}: VGGish has become a benchmark in a variety of audio recognition tasks \citep{hershey2017cnn}. We use the full pre-trained \textit{features} and \textit{embeddings} model, adding a single dropout and final linear layer to perform MC dropout for classification. We describe further modifications to the model class in Appendix \ref{sec:appendix_baseline_models}.
\end{enumerate}

\paragraph{Features}
\label{sec:benchmark:modelconf} We provide the following features for our models (see Appendix \ref{sec:appendix_feat} for details):
\begin{enumerate}

\item \textbf{Feat. A}: Features with default configuration from the VGGish \href{https://github.com/tensorflow/models/blob/master/research/audioset/vggish/vggish_input.py}{GitHub} intended for use with VGGish: 64 log-mel spectrogram coefficients using 96 feature frames of 10\,ms duration forming a single example $\mathbf{X}_i \in \mathbb{R}^{64\times 96}$ with a temporal window of 0.96 s. 

\item \textbf{Feat. B}:  Features originally designed for MozzBNNv2 (previous mosquito detection work \citep{2021pkddBNN}): 128 log-mel spectrogram coefficients with a reduced time window of 30 (from 40) feature frames and a stride of 5 frames for training. Each frame spans 64\,ms, forming a single training example $\mathbf{X}_i \in \mathbb{R}^{128\times30}$ with a temporal window of 1.92\,s. 
\end{enumerate}


\paragraph{Performance metrics}
We define the test performance with four metrics: the receiver operating characteristic area-under-curve score (ROC AUC), the precision-recall area-under-curve score (PR AUC), the true positive rate (TPR), also known as the recall, and the true negative rate (TNR), to account for class imbalances in the test sets. These are evaluated over non-overlapping feature windows of 1.92 seconds. To compare the feature sets fairly, Feat. A test data is aggregated over neighbouring windows to form decisions over 1.92\,s intervals.  Edge cases where the data cannot be partitioned into full examples are removed from the test sets.

\subsection{Task 1: Mosquito Event Detection (MED)}
\label{sec:Task1_MED}
 For mosquito event detection, we hold out Test A of labelled field data which most closely resembles the recording configuration of our system in Figure \ref{fig:humbugworkflow}. Achieving good performance on that set does not guarantee good scalability to other use cases in itself. Therefore, we also evaluate over Test B, recorded in a cage placed in a highly noisy domestic environment. As a result, the SNR is much lower than that of Test A. The statistics of the training and test sets are given in the rows of Table \ref{tab:Summary}.

\FloatBarrier
\begin{table}

\centering
\caption{\textbf{Mosquito Event Detection (MED)}. \textbf{Test A}: IHI Tanzania with HumBug Net. \textbf{Test B}: Oxford Zoology caged. Evaluated over $N_\mathrm{mozz}$ mosquito, and $N_\mathrm{noise}$ background 1.92 second samples.  30 samples drawn from each BNN to estimate the posterior. ROC AUC, PR AUC, TPR and TNR scores given as percentages ($\times 10^{2}$). The baseline ROC AUC score is given by 50 (completely random classifier). PR AUCs are relative to the prevalence of the classes, given by $N_\mathrm{mozz} / (N_\mathrm{mozz} + N_\mathrm{noise})$.}
\label{tab:results}
\centering
\footnotesize
\begin{tabular}{@{}llcccccccc@{}}
\toprule
\multirow{2}{*}{Data}   & \multirow{2}{*}{Metric} & \multicolumn{2}{c}{MozzBNNv2} & \multicolumn{2}{c}{BNN-ResNet50} & \multicolumn{2}{c}{BNN-ResNet18} & \multicolumn{2}{c}{BNN-VGGish} \\ \cmidrule(l){3-10}
                        &                         & {Feat. A }      & {Feat. B}         & {Feat. A }         & {Feat. B}          & {Feat. A }         & {Feat. B}          & {Feat. A }       & {Feat. B}        \\ \midrule
\multirow{3}{*}{\begin{tabular}[l]{@{}l@{}}\textbf{Test A}\\ \scriptsize $N_\mathrm{mozz}$: 1,714\\\scriptsize $N_\mathrm{noise}$: 2,068\end{tabular} } & ROC                     & 98.1      & 96.4     & 98.3       & 93.0        & 98.1       & 92.5         & \textbf{98.5}        & 97.3     \\
 & PR & 97.9 & 97.1 & \textbf{98.2} & 93.6 & 98.0 & 89.5 & 98.1 & 97.6 \\
                        & TPR                     & 79.5       &  79.9      & 76.9   & 79.1      & 67.0     & 76.1       & 85.6        & \textbf{87.3}    \\
                        & TNR                     & 98.3     &  98.4       & \textbf{99.0}      &  91.2        & 99.5        & 89.1    & 98.4       & 97.4       \\ \midrule
\multirow{3}{*}{\begin{tabular}[l]{@{}l@{}}\textbf{Test B}\\ \scriptsize$N_\mathrm{mozz}$: 616\\\scriptsize $N_\mathrm{noise}$: 1,084\end{tabular}}   & ROC                     & 71.1      & 58.4          & 74.8   & 76.1            & 71.1         & \textbf{77.0}        & 74.1        & 57.4      \\
& PR & 64.0 & 63.2 & 72.0 & \textbf{75.0} & 68.5 & 74.9 & 70.7 & 61.3 \\ 
                        & TPR                     & 30.1       & 30.9        & 31.0 & \textbf{34.1}               & 30.6        & 32.8      & 30.8        &   31.7  \\
                        & TNR                     & 99.3      & 99.2       & \textbf{100.0} & 98.8           & \textbf{100.0}         & 99.3         & \textbf{100.0}       & 99.3     \\ \bottomrule
\end{tabular}
\end{table}

For the intended use case of Test A, all of the model and feature combinations were able to achieve ROC AUC above 0.93 and PR AUC above 0.90 (Table \ref{tab:results}). Furthermore, all of the models improve in performance when utilising Feat. A  over Feat. B. 
However, performance on Test B is significantly lower for all models with no clear preference for features. The highest AUCs are achieved by BNN-ResNet when trained on Feat. B (ResNet18: ROC: 0.770, PR: 0.749, ResNet50: ROC: 0.76, PR: 0.750). To verify that the issue does not lie in the test set, after manually verifying each label resulting from feature extraction, we trained a model on half of Test B to achieve an ROC AUC of 0.915 on the second half of Test B. (Appendix \ref{sec:appendix_testPerformance}, Figure \ref{fig:TestResultsB}). Furthermore, prior work was able to achieve ROC AUCs of 0.871 to 0.952 
with smaller neural networks which were optimised for use with scarce data \citep{kiskin2017mosquito}. The task presented in this paper, however, is to be able to achieve good performance over Test B, in addition to Test A, without the model having access to any data (or covariates) from either test set during training. This task therefore poses a challenge to promote the development of generalisable deep learning models, which we require for robust deployment.

\subsection{Task 2: Mosquito Species Classification (MSC)}
\label{sec:Task2_Species}



This task utilises data collected with a wide range of well-populated species of wild captured mosquitoes at IHI Tanzania. 
We split the 8 most populated species by recordings (each \texttt{audio\_id} records a unique mosquito)\ikn{This is NOT the case with all the datasets. I should either make that clear of think of upgrading the metadata storage to capture this}  into a 75-25\,\% train-test partition through a range of 5 fixed random seeds. To address data imbalance, upon training, we supply class weights as the inverse of the class frequency. From our experiments, this strategy has produced better results versus downsampling majority or oversampling minority classes, but there is likely room for improvement to be found here with paradigms such as few-shot learning \citep{sun2019meta}, loss-calibrated inference \citep{cobb2018loss}, and many more.
To further motivate our two-stage pipeline, we note that the start and stop time tags for this dataset were auto-generated with a prior BCNN \citep{2021pkddBNN}. These factors contribute to a realistic test-bed for our pipeline of Figure \ref{fig:humbugworkflow}, and hence any models developed for this dataset are candidates for real-world deployment.

\begin{table}[ht]
\caption{\textbf{Mosquito Species Classification (MSC):} Statistics, ROC AUC and PR AUC scores on the cup recordings conducted at IHI Tanzania. The total AUCs are given by the micro average. The baseline ROC AUC score is given by 50 (completely random classifier). PR AUC scores are relative to the prevalence of the classes, given by the number of (test) mosquitoes per class divided by the total number of mosquitoes (test). All scores are reported as mean (standard deviation) over 5 random train-test partitions ($\times 10^{2}$) of unique wild \textit{`mosquitoes'}, with the distribution of column 1 in the form of train (test), prevalence (\%). }
\label{tab:multispecies}
\centering
\scriptsize
\begin{tabular}{@{}llcccccccc@{}}
\toprule
\multirow{3}{*}{\begin{tabular}[l]{@{}l@{}}\textbf{\textit{Mosquito}}\\Train (test),\\Prevalence\end{tabular}} & \multirow{2}{*}{Metric}  & \multicolumn{2}{c}{MozzBNNv2} & \multicolumn{2}{c}{BNN-ResNet50} & \multicolumn{2}{c}{BNN-ResNet18} & \multicolumn{2}{c}{BNN-VGGish} \\ \cmidrule(l){3-10} 
                      &   &  Feat. A        &  Feat. B        &  Feat. A          & Feat. B          &  Feat. A         & Feat. B          &  Feat. A          & Feat. B      \\ \midrule
\rowcolor{Gray} \textbf{\textit{An. arabiensis}}       & ROC        &     83.7 (1.2) &     \textbf{86.6 (1.0)} &    75.8 (7.3) &    84.9 (2.4) &    75.6 (7.7) &    83.4 (8.7) &   85.7 (2.2) &  84.1 (1.5) \\
385 (129), 36\%    &  PR &   77.5 (2.5) &     \textbf{80.9 (1.6)} &    71.8 (5.8) &    80.3 (4.4) &    67.9 (9.7) &    78.5 (8.8) &   80.2 (3.9) &  77.3 (2.2) \\
\rowcolor{Gray}\textbf{\textit{Culex pipiens}}     &    ROC  &     81.4 (1.2) &     \textbf{86.7 (1.4)} &    85.0 (2.2) &    84.0 (3.3) &    85.0 (2.5) &    85.6 (4.8) &   82.1 (1.7) &  81.4 (1.6) \\
252 (84),  24\%     & PR     &     57.3 (3.3) &     66.9 (2.3) &    61.4 (4.4) &    60.1 (5.6) &    60.3 (7.6) &    \textbf{67.6 (8.3)} &   59.0 (3.6) &  59.0 (3.0) \\
\rowcolor{Gray}\textbf{\textit{Ae. aegypti}}  &  ROC &   95.0 (0.8) &     96.4 (1.9) &    \textbf{98.8 (0.6)} &    97.1 (1.8) &    98.2 (0.3) &    94.5 (1.1) &   96.6 (1.0) &  96.3 (2.3) \\
36 (13), 3.6\%  & PR     &     53.8 (7.2) &     74.4 (5.1) &    \textbf{83.0 (2.7)} &   78.0 (11) &    76.6 (3.9) &    75.9 (3.1) &   66.6 (7.7) &  76.0 (4.9) \\
\rowcolor{Gray}\textbf{\textit{An. funestus ss}} & ROC   &     91.7 (0.6) &     92.3 (1.3) &    \textbf{93.8 (2.1)} &    84.7 (7.2) &    85.5 (7.7) &    90.6 (4.9) &   93.5 (1.4) &  91.0 (1.5) \\
186 (62), 17.5\%  &  PR     &     78.2 (1.9) &     80.9 (1.1) &    \textbf{84.6 (4.5)} &   70.9 (10) &   67.2 (14) &    77.4 (9.6) &   83.3 (3.3) &  76.0 (4.2) \\
\rowcolor{Gray}\textbf{\textit{An. squamosus}}    &   ROC     &     78.2 (1.9) &     85.2 (2.4) &    \textbf{88.8 (4.4)} &    85.2 (5.3) &    86.5 (3.2) &    83.5 (3.9) &   83.6 (3.3) &  86.4 (2.9) \\
68 (23), 6.5\%   & PR       &     21.1 (3.3) &     35.6 (5.8) &   39.4 (10) &    34.5 (8.5) &    36.0 (6.2) &    \textbf{40.3 (9.8)} &   28.6 (8.1) &  35.6 (6.1) \\
\rowcolor{Gray}\textbf{\textit{An. coustani}}    & ROC  &     90.8 (2.3) &     88.4 (3.2) &    \textbf{93.4 (1.4)} &    85.1 (4.6) &    92.2 (2.3) &    83.6 (5.5) &   89.9 (4.6) &  85.2 (4.1) \\
37 (13), 3.6\% & PR     &     32.7 (8.0) &     26.6 (8.4) &    \textbf{35.2 (8.5)} &   23.4 (11) &   32.5 (16) &    26.4 (9.8) &  33.2 (10) &  25.7 (8.2) \\
\rowcolor{Gray}\textbf{\textit{Ma. uniformis}}    & ROC     &     82.5 (7.6) &     82.0 (6.4) &    \textbf{84.7 (6.9)} &    83.6 (9.4) &    87.5 (4.5) &    80.1 (8.8) &   83.4 (2.2) &  77.2 (8.3) \\
57 (19), 5.4\%  &  PR     &     33.9 (8.7) &     29.6 (9.0) &   35.4 (10) &   34.5 (13) &    \textbf{35.9 (7.8)} &   35.4 (13) &   29.1 (4.5) &  23.4 (5.2) \\
\rowcolor{Gray}\textbf{\textit{Ma. africanus}} & ROC  &     91.2 (3.0) &     91.3 (1.7) &    \textbf{93.0 (2.4)} &    84.5 (8.9) &    89.9 (4.6) &    85.8 (4.3) &   92.0 (2.6) &  91.1 (2.2) \\
28 (10), 2.8\% & PR   &     26.8 (9.7) &     22.3 (5.0) &   29.0 (10) &   22.7 (19) &   24.3 (11) &    21.9 (4.2) &   \textbf{33.5 (8.8)} &  23.4 (3.2) \\\midrule
\rowcolor{Gray}\textbf{Total} &  ROC  &    91.4 (0.8) &     \textbf{92.7 (0.9)} &    89.9 (2.5) &    90.4 (2.1) &    90.1 (2.1) &    90.8 (3.1) &   92.1 (1.2) &  91.4 (0.7) \\ 
1049 (353)   &  PR   &   66.9 (2.1) &     \textbf{71.6 (2.2)} &    63.4 (4.8) &    65.0 (3.8) &    57.7 (7.3) &    69.2 (8.4) &   68.1 (3.9) &  66.2 (2.0) \\

\bottomrule
\end{tabular}
\end{table}

The ROC AUC of 0.927 and PR AUC of 0.716 produced for this classification problem (Table \ref{tab:multispecies}) by the best-performing baseline model, MozzBNNv2-FeatB, demonstrate the ability to discriminate between different species of mosquitoes that have been sampled individually in the wild. 

The results also show how our dataset is well suited for training multi-species classifiers to a degree that was not available previously. From the total ROC and PR AUCs, there is a slight preference for Feat. B for all models, except VGGish (as Feat. A were naturally made to be used with the model). 

When interpreting PR AUC scores, a good indication of model performance is given by the increase in PR AUC over the baseline prevalence, given in the first column of Table \ref{tab:multispecies}. Due to the heavy class imbalance, the PR AUC scores are significantly lower on the minority classes, except for \textit{Ae. aegypti} mosquitoes, which may be due to their larger size and hence more distinct difference in acoustic properties. The model confusion occurs in species with similar physical characteristics (see Appendix \ref{sec:appendix_testPerformance}, Figure \ref{fig:spec_multispecies} for a visualisation of spectra for each species).  Example class-specific softmax outputs, ROC and PR curves, as well as confusion matrices are discussed in further detail in Appendix \ref{sec:appendix_testPerformance}.

Maximising PR performance of the under-represented, lower-scoring, classes, is the primary area in need of improvement in this task, which we encourage researchers to explore further.

\FloatBarrier
\section{Conclusion}
\label{sec:conclusion}
In this paper we present a database of 20 hours of finely labelled mosquito sounds and 15 hours of associated non-mosquito control data, constructed from carefully defined recording paradigms. Our recordings capture a diverse mixture of 36 species of mosquitoes from controlled conditions in laboratory cultures, as well as mosquitoes captured in the wild. The dataset is a result of a global co-ordination as part of the HumBug project.
Our paper makes the significant contribution of providing both the large multi-species dataset and the infrastructure surrounding it, designed to make it straightforward for researchers to experiment with.


Despite decades of work, mosquito-borne diseases are still dangerous and prevalent, with malaria alone contributing to hundreds of thousands of death each year. Therefore a further contribution of this work is to make available mosquito data that is still a scarce commodity. In addition, we have highlighted that our dataset contains real field data collected from smartphones, as well as varying background environments and different experimental settings. As a result, this multi-species data set will continue to help domain-experts in the bio-sciences study the spread of mosquito-carrying diseases, as well as the myriad of factors that affect acoustic flight tone.

Finally, HumBugDB will be of interest to machine learning researchers working with acoustic data, both in the challenges posed by real-world acoustic data, as well as in the way that we use Bayesian neural networks for mosquito event detection and species classification. We provide baseline models alongside extensive documentation. As a result, we make it easy for researchers to start building their own models.
It is our aim, by releasing this dataset and identifying areas for improvement in our baseline tasks, to encourage further work in the detection of mosquitoes. We hope this in turn leads to improved future detection and classification algorithms.

\begin{ack}
This work has been funded from a 2014 Google Impact Challenge Award, and has received support from the Bill and Melinda Gates Foundation, [\#opp1209888] since 2019. We would like to thank Paul I Howell and Dustin Miller (Centers for Disease Control
and Prevention, Atlanta), Dr. Sheila Ogoma (The United
States Army Medical Research Unit in Kenya (USAMRU-K)).
Prof. Gay Gibson (Natural Resources Institute, University of
Greenwich) and Dr. Vanessa Chen-Hussey and James Pearce
at the London School of Tropical Medicine and Hygiene. For
significant help and use of their field site Prof. Theeraphap
Chareonviriyaphap and members of his lab, specifically Dr.
Rungarun Tisgratog and Jirod Nararak (Dept of Entomology,
Kasesart University, Bangkok) and Dr. Michael J. Bangs
(Public Health \& Malaria Control International SOS Kuala
Kencana, Papua, Indonesia). We also thank nVIDIA for the grant of a Titan
Xp GPU.

\end{ack}

\bibliography{MozzBNNNeurIPS}

%% file: neurips_data_2021_supplement.tex
\newcolumntype{g}{>{\columncolor{Gray}}c}



\appendix

\section*{HumBugDB: supplementary materials} 
The supplementary materials include:
\begin{itemize} \item  Code and data licensing in Section \ref{sec:appendix_license}.
\item A code manual with additional discussions for the results of the main paper in Section \ref{sec:appendix_code_use}.
\item Section \ref{sec:appendix_db_metadata} supplies details on the database schema, and provides an explanation for every field of the metadata.
\item The datasheet for HumBugDB is given in Section \ref{sec:appendix_datasheet_for_dataset}.
\end{itemize}
\section{Licenses}
\label{sec:appendix_license}
\subsection{Code license}
MIT License

Copyright (c) 2021 HumBug-Mosquito

Permission is hereby granted, free of charge, to any person obtaining a copy
of this software and associated documentation files (the "Software"), to deal
in the Software without restriction, including without limitation the rights
to use, copy, modify, merge, publish, distribute, sublicense, and/or sell
copies of the Software, and to permit persons to whom the Software is
furnished to do so, subject to the following conditions:

The above copyright notice and this permission notice shall be included in all
copies or substantial portions of the Software.

THE SOFTWARE IS PROVIDED "AS IS", WITHOUT WARRANTY OF ANY KIND, EXPRESS OR
IMPLIED, INCLUDING BUT NOT LIMITED TO THE WARRANTIES OF MERCHANTABILITY,
FITNESS FOR A PARTICULAR PURPOSE AND NONINFRINGEMENT. IN NO EVENT SHALL THE
AUTHORS OR COPYRIGHT HOLDERS BE LIABLE FOR ANY CLAIM, DAMAGES OR OTHER
LIABILITY, WHETHER IN AN ACTION OF CONTRACT, TORT OR OTHERWISE, ARISING FROM,
OUT OF OR IN CONNECTION WITH THE SOFTWARE OR THE USE OR OTHER DEALINGS IN THE
SOFTWARE.

\subsection{Database license}
CC-BY-4.0, \url{https://creativecommons.org/licenses/by/4.0/}

\clearpage
\section{Code use}
\label{sec:appendix_code_use}
\subsection{Code access and structure}
\label{sec:appendix_code_access}

\begin{itemize}
    \item The audio recordings and metadata \texttt{csv} are hosted on Zenodo \url{http://doi.org/10.5281/zenodo.4904800}. under a CC-BY-4.0 license. 
    \item Code (and the metadata \texttt{csv} for completeness) is hosted on \url{https://github.com/HumBug-Mosquito/HumBugDB} under the MIT license.
\end{itemize}

The GitHub data directory structure as of commit \href{https://github.com/HumBug-Mosquito/HumBugDB/tree/50656758594982480f568598874f79c222432e01}{50656758594982480f568598874f79c222432e01} is as follows:
\\
\dirtree{%
.1 HumBugDB.
.2 README.md.
.2 *requirements.txt.
.2 notebooks.
.3 main.ipynb.
.3 species\_classification.ipynb.
.3 supplement.ipynb.
.3 spec\_audio\_multispecies.ipynb.
.2 data.
.3 metadata.
.4 *.csv.
.3 audio.
.4 *.wav.
.2 lib.
.3 PyTorch.
.4 \_\_init\_\_.py.
.4 vggish.
.4 ResNetDropoutSource.py.
.4 ResNetSource.py.
.4 runTorch.py.
.4 runTorchMultiClass.py.
.3 Keras.
.4 \_\_init\_\_.py.
.4 config\_keras.py.
.4 runKeras.py.
.3 config.py.
.3 feat\_vggish.py.
.3 feat\_util.py.
.3 evaluate.py.
.3 write\_audio.py.
.2 outputs.
.3 models.
.4 keras.
.4 pytorch.
.3 features.
.3 plots.
}

A \texttt{README} and several \texttt{requirements} are included for installing Keras, PyTorch, and dependencies for the code. 
The metadata is located in \texttt{/data/metadata/} as a \texttt{csv} file.

Extract the audio from Zenodo to the folder \texttt{/data/audio/} and launch the Jupyter notebook \texttt{main.ipynb} to perform train-test splitting, feature extraction, model training, and evaluation for Task MED: mosquito event detection. The notebook imports from \texttt{lib} the necessary files depending on the choice of kernel and PyTorch or Keras. Task MSC: mosquito species classification is addressed in \texttt{species\_classification.ipynb}. Remaining supplementary material is found in \texttt{supplement.ipynb} and \texttt{spec\_audio\_multispecies.ipynb}.

\subsection{Code manual}
\label{sec:appendix_code_manual}
\paragraph{Overview}
The following documentation has last been verified with the \href{https://github.com/HumBug-Mosquito/HumBugDB/tree/50656758594982480f568598874f79c222432e01}{commit 50656758594982480f568598874f79c222432e01}. Future code aims to maintain compatibility where possible. However, please visit the GitHub repository \url{https://github.com/HumBug-Mosquito/HumBugDB} for the most comprehensive instructions and updates. Latest development code can be found on the \texttt{devel} branch, and the stable version on \texttt{master}. Releases with large binaries (any pre-trained models or features) can be found on \url{https://github.com/HumBug-Mosquito/HumBugDB/releases}. 

\paragraph{Top-level notebook (MED)} \texttt{main.ipynb} performs data partitioning,  feature extraction and segmentation in \texttt{get\_train\_test\_from\_df()}, model training in \texttt{train\_model()}, and model evaluation in \texttt{get\_results()}. The code is configured with \texttt{config.py}, where data directories are specified for the data, metadata and outputs, and feature transformation parameters are supplied. Model hyperparameters are given in \texttt{config\_keras.py} or \texttt{config\_pytorch.py}. The notebook supports both Keras \citep{chollet2015keras} and PyTorch \citep{NEURIPS2019_9015} with a common interface for convenience. In more detail, each top-level function is described as follows:
\begin{itemize} \item \texttt{get\_train\_test\_from\_df(df\_train, df\_test\_A, df\_test\_B)}
extracts, reshapes, strides, and normalises features for use as tensors, and saves them to \texttt{config.dir\_out}, if features with that particular configuration do not exist already. This function supports the creation of features for any feature-model combination. If one wishes to extract Feat. A, the function is imported from \texttt{feat\_vggish.py}. For Feat. B, \texttt{get\_train\_test\_from\_df()} is imported from \texttt{feat\_util.py}. The choice of import is specified in the notebook cells. Section \ref{sec:appendix_feat} discusses the features in more depth. 

The data is split into train and test based on the matches of experiment ID to the audio tracks from the metadata given in \texttt{df\_train, df\_test\_A, df\_test\_B}. It is important that no test recordings from these experiments are seen during training in advance, as otherwise model performance is overestimated.  

\item \texttt{train\_model(X\_train, y\_train, X\_val=None, Y\_val=None, model=ResnetDropoutFull())} trains the BNNs on the data supplied (with validation data optional). The assumed input shape is that of the features produced by \texttt{get\_train\_test\_from\_df()}. The \texttt{model} argument is optional, and can take any model class defined in \texttt{runTorch.py}. The model architecture and training strategies may be changed further in \texttt{runKeras.py} or \texttt{runTorch.py}.

\item \texttt{get\_results(model, X, y, filename, n\_samples=1)} evaluates the model object on test data \texttt{\{X, y\}} with the number of MC dropout samples as \texttt{n\_samples}. If using deterministic networks, leaving the input argument blank will default to a single evaluation. For any option using Feat. A, the output is aggregated over neighbouring windows to produce predictions over the same window size as Feat. B for fair benchmarking. If you wish to use raw windows (e.g. for creating precise start/stop tags), you may modify this behaviour by removing \texttt{resize\_window()} from \texttt{evaluate.py}. Specify the output plot directory in \texttt{config.plot\_dir}, and the output filename in \texttt{filename}.
\end{itemize}

{\paragraph{Species classification notebook (MSC)} \texttt{species\_classification.ipynb} is used to classify mosquito species on a subset of data from the database. The underlying functions for feature creation and model training are shared as much as possible from the same library sources. Some alterations have been performed for support with PyTorch multi-class classification, due to a difference in API for certain loss functions (BCE loss vs XEnt loss of binary vs multi-class problems).
\begin{itemize}
    \item \texttt{get\_feat\_multispecies(df\_all, train\_fraction, random\_seed)} extracts features. Configuration for feature extraction is given in \texttt{config.py}. The fraction of data used for training is given by \texttt{train\_fraction}, for which we use 0.75 for MSC. The random seeds used are [5, 10, 21, 42, 100]. The outputs are returned in either list or tensor form, depending on whether features require aggregation (Feat A.) or not (Feat B.). Outputs shapes are designed to work without re-shaping for the evaluation function \texttt{get\_results\_multiclass().}
    \item \texttt{train\_model()} follows the same input and output structure as \texttt{main.ipynb}. Models are defined and selected from \texttt{runTorchMultiClass.py} or \texttt{runKeras.py}.
    \item \texttt{get\_results\_multiclass()} produces outputs of ROC-AUC scores per class with class averages, and confusion matrices in \texttt{.pdf} and \texttt{.txt} form. As before, the plot directory is specified in \texttt{config.py}. All outputs are given over 1.92 second windows.
\end{itemize}

\paragraph{Supplementary notebook} \texttt{supplement.ipynb} is used to reproduce the plots of species distribution in this paper (Figure \ref{fig:species_dist}) and contains utilities that were used for debugging and visualising the data, should they be helpful for researchers using their own functions.

\paragraph{Spec audio multispecies notebook} \texttt{spec\_audio\_multispecies.ipynb} is used to create the plots of Figure \ref{fig:spec_multispecies} and contains utilities for visualising mosquito samples of any species.

\subsection{Feature parameters}
\label{sec:appendix_feat}
We first need to define the number of feature windows that are used to represent a sample, $\mathbf{X}_i \in \mathbb{R}^{h\times w}$, where $h$ is the height of the two-dimensional matrix, and $w$ is the width. The longer the window, $w$, the better potential the network has of learning appropriate dynamics, but the smaller the resulting dataset in number of samples. It may also be more difficult to learn the salient parts of the sample that are responsible for the signal, resulting in a weak labelling problem \citep{kiskin2019super}. Early mosquito detection efforts have used small windows due to a restriction in dataset size. For example, \citet{PotamitisKaggle} supplies a rich database of audio, however the samples are limited to just under a second. However, despite the mosquito's simple harmonic structure, its characteristic sound also derives from the temporal variations, as is visible from spectrograms.  We suspect this flight behaviour tone is better captured over longer windows, however we encourage researchers to experiment, for example by padding with noise to match the window size of this architecture, or by choosing a smaller window to extract features from. 

The features used in the  MED and MSC tasks are as follows:
\begin{enumerate}
\item \textbf{Feat. A}: Features with default configuration from the VGGish \href{https://github.com/tensorflow/models/blob/master/research/audioset/vggish/vggish_input.py}{GitHub} intended for use with VGGish: $\mathbf{X}_i \in \mathbb{R}^{64\times 96}$, 64 log-mel spectrogram coefficients using 96 feature frames of 10\,ms duration. To compare feature sets fairly, predictions are aggregated over neighbouring windows to create outputs over 1.92 second windows as used in Feat. B.
\item \textbf{Feat. B}:  Features of previous acoustic mosquito detection work \citep{2021pkddBNN}: 128 log-mel spectrogram coefficients with a reduced time window of 30 (from 40) feature frames and a stride of 5 frames for training. Each frame spans 64\,ms, forming a single training example $\mathbf{X}_i \in \mathbb{R}^{128\times30}$ with a temporal window of 1.92\,s. To create an augmented dataset, we stride the input signal feature window with a step of 5 feature windows (a duration of 320\,ms) Note that the training data is segmented by using overlapping strides specified with \texttt{config.step\_size=5}, whereas the test data is created with no overlap. Samples that do not divide evenly into the window size are discarded (this is a very small number when using such a small step, and we prefer this option over padding with zeros or noise, though alternate solutions are welcome).
\end{enumerate}

Detailed parameterisation is supplied in Table \ref{tab:Featureparams}.

\begin{table}[ht]
\centering
\caption{Feature transformation parameters, in samples unless otherwise indicated. Audio processed with \texttt{librosa} for Feat. B and its own implementation in VGGish. The size of 1 frame in $w$ is equal to \texttt{hop\_length}. For the parameterisation of Feat. B this is 64 ms, resulting in an input feature slice of $512/8000 \times 30 = 1.92$ s duration and $h=128$ height. For Feat. A, the example window width is $1600/16000 \times 96 = 0.96$ s with height $h=64$ (log-mel coefficients). Feat. A parameters calculated from \texttt{mel\_features.py} with the default values supplied in \texttt{vggish\_params.py}.}
    \label{tab:Featureparams}
\begin{tabular}{@{}llllllll@{}}
\toprule
Method & Sample rate & \texttt{NFFT} & \texttt{win\_size} & \texttt{hop\_length}  & $h$ (\texttt{n\_mels}) & $w$ (frames)  & Stride  \\ \midrule
Feat. A & 16,000 & 512 & 400  & 160 & 64 & 96 & 160 \\ 
Feat. B & 8,000 & 2,048 & 2,048      & 512  & 128     & 30  & 512  \\ 
\bottomrule
\end{tabular}
\end{table}

From the results of Section \ref{sec:benchmark}, it appears models are able to learn better representations with Feat. B over Feat. A for the purpose of MSC. However, Feat. A achieve better results in the MED task. We encourage further study and window size experimentation for determining if there is an optimum window size for any given task.


\subsection{Baseline models}
\label{sec:appendix_baseline_models}

\paragraph{MozzBNNv2}
We give the full model structure in Figure \ref{fig:BCNN}. Lambda layers are dropout layers which are placed to perform MC dropout at test-time. This structure bares similarity to VGGish\footnote{\url{https://github.com/tensorflow/models/tree/master/research/audioset/vggish}}, which uses 0.96 second log-mel spectrogram patches as inputs, and 11 weight layers (primarily convolutional layers and max-pool layers). Furthermore, this is an incremental improvement over the model used in \citet{2021pkddBNN}, \url{https://github.com/HumBug-Mosquito/MozzBNN}. To ensure this model does not have an advantage in benchmark tasks, each model structure was re-trained with the same data as detailed throughout the main text.
\begin{figure}

    \centering
    \includegraphics[height=0.93\textheight]{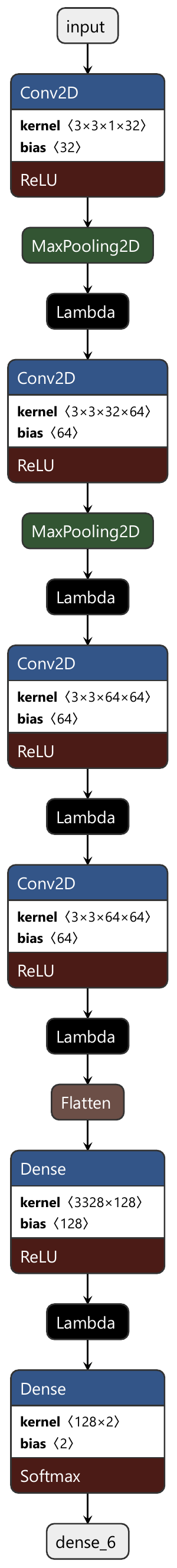}
    
    \caption{BCNN Keras model. Log-mel spectrograms are input with $w=30, h=128$, and passed through the above model. Lambda layers are dropout layers with probability 0.2. Made with \url{https://github.com/lutzroeder/netron}.}
    
    \label{fig:BCNN}
\end{figure}

\paragraph{PyTorch ResNet-X}
 We modify the final layers for compatibility with our data (in \texttt{runTorch.py}). Furthermore, we have augmented the construction blocks \texttt{BasicBlock()} and \texttt{Bottleneck()}, as well as the overall model construction, to feature dropout layers to act as an approximation for the model posterior at test-time. Dropout is implemented implicitly in \texttt{ResNetSource.py}, to not interfere with the behaviour of \texttt{model.eval()}, which by default disables dropout layers at test-time, removing the necessary stochastic component. We have pre-defined two configurations as \texttt{Resnet50DropoutFull} and \texttt{Resnet18DropoutFull}, which are both passed as objects to input arguments of model training or loading code. For further modifications see \texttt{runTorch.py} for MED and \texttt{runTorchMultiClass.py} for MSC. For ResNet-18, and ResNet-34 the final \texttt{self.fc1} layer is of size \texttt{[512,N]}, whereas for ResNet-50 the size is \texttt{[2048,N]}, where \texttt{N} is the number of classes for the cross-entropy loss function, or 1 if used with the binary cross-entropy loss. A quick way to check the requirement is to print \texttt{x.shape()} before the creation of the \texttt{fc1} layer.

\paragraph{PyTorch VGGish}
The source code for this model can be found at \url{https://github.com/harritaylor/torchvggish}, which is a PyTorch port of \url{https://github.com/tensorflow/models/tree/master/research/audioset/vggish}. The adaptation of this model class to transfer learning problems is open to interpretation. We opt to use the most straightforward method, which is to connect the output of the \textit{embeddings} layer to a linear layer which then feeds into a sigmoid and binary cross-entropy loss function (in \texttt{runTorch.py}, or straight into a categorical cross-entropy loss function in \texttt{runTorchMultiClass.py}. We then re-train the network, with the weights pre-trained on AudioSet. Dropout is used in the last layer of this model only, to create a pseudo-BNN which has estimates of uncertainty when sampled at test time. 

With native features (Feat. A), no further modifications were made, though we note that normalising the output of the embedding layer (division by 255) before connecting to a final linear layer provided a boost to performance that may be beneficial to the model overall. As described, this model is implemented as class \texttt{VGGishDropout(nn.Module)}. 

When utilising Feat. B, the size of the output changes and thus a slight tweak to the final layers in the \textit{embeddings} module was required: this involved removing layer (0) in \textit{embeddings} as the dimension of \texttt{in\_features} changed from 12,288 to 4,096.

We also note, finally, that this network proved troublesome with high (or sometimes default) values of learning rate for the Adam optimiser which we used for the PyTorch training loop. We lowered the learning rate from 0.0015 to 0.0003 which alleviated this problem, but it is worth keeping in mind that the learning rate should be tailored to the type of model and criterion utilised (in \texttt{config\_pytorch.py}).

\paragraph{Model training}
To select the loss that is used to define the best performing model, edit \texttt{runTorch.py} to make use of \texttt{train\_acc} (or any other metric as desired) by replacing. Similarly, amend the training epoch loop to change other metrics or properties during training. In \texttt{runKeras.py}, supply arguments and any other desired callbacks and model checkpointing strategies to \texttt{model.fit()}. For all models in the MED task, the validation accuracy  on a random split of the training data has been used to checkpoint the best-performing model. 

 For MSC, no validation sets were used during the training of the models due to the way the data was partitioned and general data scarcity per class. The loss function for PyTorch models was also changed from binary cross-entropy (\url{https://pytorch.org/docs/stable/generated/torch.nn.BCELoss.html}) to categorical cross-entropy (\url{https://pytorch.org/docs/stable/generated/torch.nn.CrossEntropyLoss.html}). \textbf{Be aware} that \texttt{BCELoss} \textbf{does not include} a sigmoid/softmax layer before input to the loss function, which resulted in the creation of separate models for the binary and multi-class classification problems. The models are therefore stored separately in \texttt{runTorch.py} and \texttt{runTorchMultiClass.py} for the binary (MED) and multi-class (MSC) problems respectively. The Keras model remained unaffected and shares the exact same training code between the MSC and MED tasks.

\paragraph{Hardware} The code was developed on Ubuntu 20.04 with an i7-8700K CPU, 32\,GB RAM and a Titan Xp GPU with 12\,GB VRAM, but models were trained and optimised with lower end hardware (Windows 10, Intel i7-4790K CPU with 16\,GB RAM and a GTX970 GPU with 4\,GB VRAM).

\paragraph{Memory optimisation}
Note that the default settings require at least 16\,GB RAM to load into memory for ResNet-50 processing, as channels are replicated 3 times to match the pre-trained weights model. To reduce the strain on memory, increase the \texttt{step\_size} parameter in \texttt{config.py} to reduce the number of windows created by feature extraction. This reduces the overlap between samples. 

Alternatively, it is possible to use a non-pretrained architecture and change the tensor creation code in \texttt{build\_dataloader()} from \texttt{runTorch.py} to remove \texttt{.repeat(1,3,1,1)} as there will be no need to copy over identical data over three channels. 

Note that once the tensors have been created, VRAM is not an issue due to the batching over the dataloaders (this code has been run on a GTX970 with 3.5\,GB useable VRAM).

A further alternative is creating dataloaders which stream either the raw files and create features on the fly, or stream pre-stored features.

\paragraph{Hyperparameters} Configure the hyperparameters in \texttt{config\_pytorch.py} and \texttt{config\_keras.py}. The number of epochs was set by observing the learning rate of the network. For MED, within a few epochs, the models began to strongly overfit, with the training accuracy failing to improve validation accuracy. For this reason, both models are set to a low epoch number, and have a fairly low \texttt{max\_overrun} counter, which determines the maximum number of steps taken for which the target metric fails to improve. The dropout rate and batch size were set to 0.2 and 32, values which are generally risk-free. We note here that the point at which we stop training the model made a fairly significant difference to the balance between true positive and true negative errors (despite a similar overall ROC AUC score). In this respect, the optimisation procedure for the models could be improved with more careful thought about the metrics used for training. If error types are important, consider using loss-calibrated approaches such as that of \citet{cobb2018loss}.

For MSC, the learning rates are reduced and number of epochs significantly increased. The models all took many more epochs before training stalled. There is also an advantage to using a larger batch size e.g. 128 (and thus higher probability of encountering a greater number of classes per epoch) for faster convergence if desired.

\subsection{Test performance}
\label{sec:appendix_testPerformance}
\subsubsection{Verifying data integrity of Test B}
To support the validity of Test B: the cage recordings conducted at Oxford Zoology, we train the Keras model on half of Test B and test on the other half (with recordings held out), with the settings: \texttt{epochs = 7, tau = 1.0, dropout = 0.2, validation\_split = None, batch\_size = 32, lengthscale = 0.01} to achieve the results of Figure \ref{fig:TestResultsB}. Figure \ref{fig:testB_MI} illustrates the raw output of the mean model probability across 10 MC dropout samples, alongside the predictive entropy and mutual information. The ground truth is given in red, dotted. The model produces a respectible ROC of 0.915, far outperforming the score it achieves when not trained on any part of this dataset (of 0.770). There is therefore a property of this dataset which is not captured well within the rest of the training data (perhaps the SNR or type of background noise encountered), which warrants further research.

\begin{figure}
\centering
\begin{subfigure}[t]{0.35\textwidth}
\centering
      \includegraphics[width=1.0\textwidth]{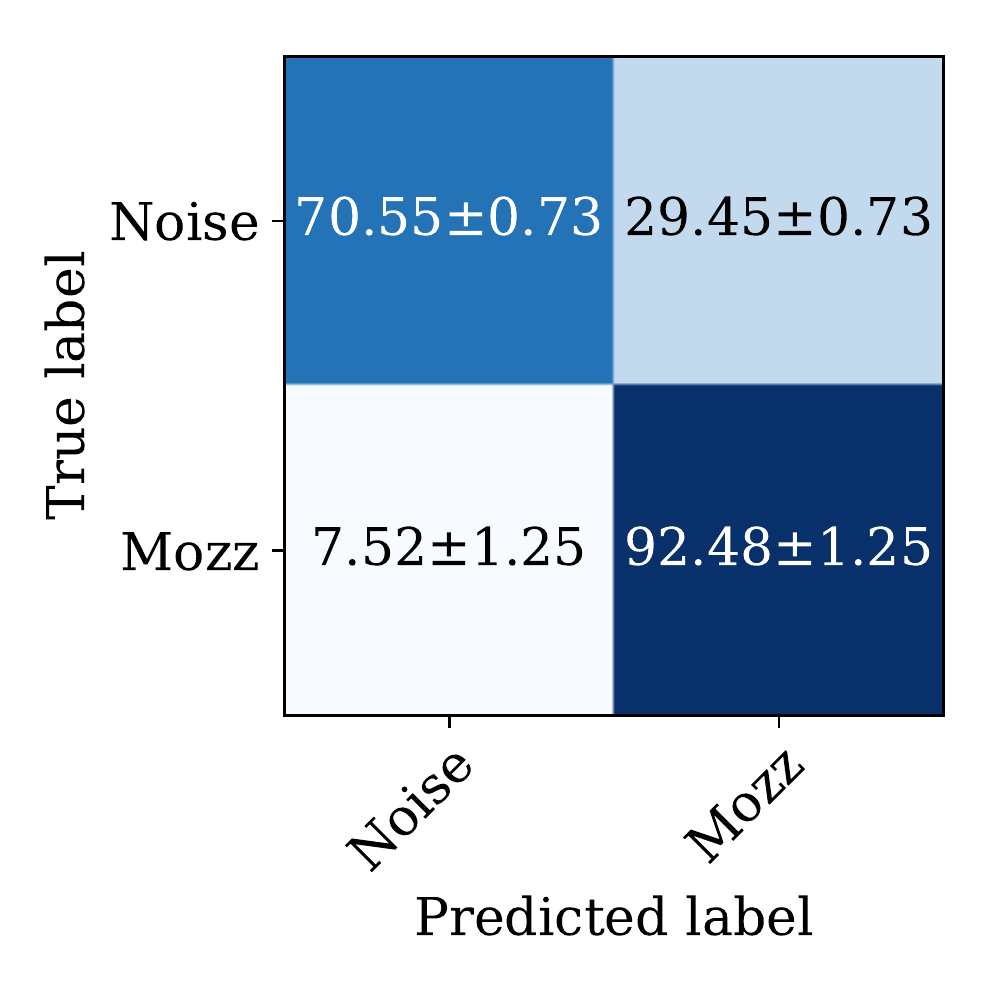}
    \caption{Confusion matrix}
    \label{fig:testB_cm}
    \end{subfigure}
    \begin{subfigure}[t]{0.35\textwidth}
    \centering
    \includegraphics[width=1.0\textwidth]{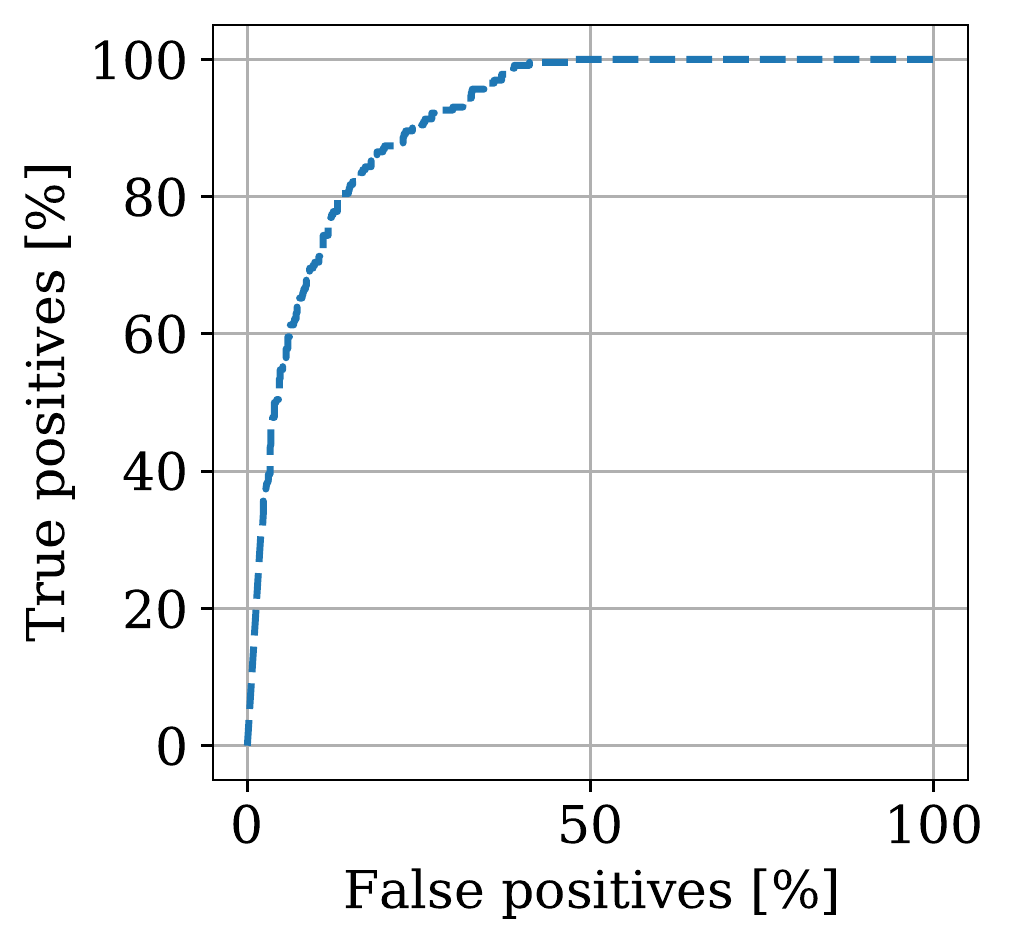}
    \caption{ROC AUC: $0.915 \pm 0.003$}
    \label{fig:testB_ROC}
\end{subfigure}

 \begin{subfigure}[t]{0.9\textwidth}
    \centering
    \includegraphics[width=1.0\textwidth]{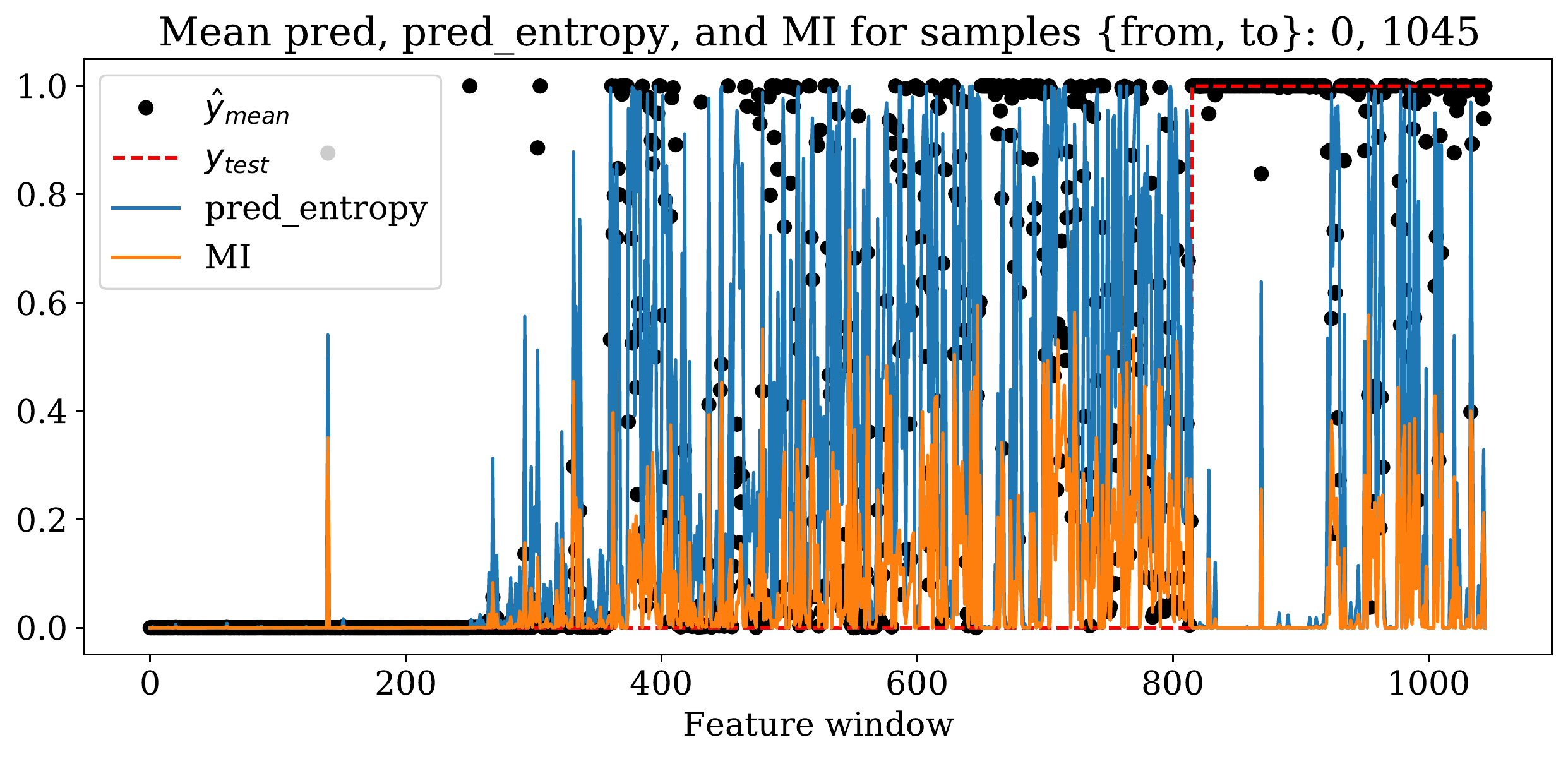}
    \caption{Mean prediction, predictive entropy and mutual information for feature windows}
    \label{fig:testB_MI}
\end{subfigure}
    \caption{Performance on half of a hold-out test set constructed from Test B. Confusion matrices given in normalised percentage, and ROC in the form of mean $\pm$ standard deviation, across $N=10$ MC dropout samples.}
    \label{fig:TestResultsB}
\end{figure}

\FloatBarrier

\subsubsection{Mosquito Species Classification (MSC)} 
This section continues the discussion of \ref{sec:Task2_Species}. 
To illustrate typical model performance, we will take a single random seed from the best-performing model, MozzBNNv2, trained on Features B. We show the ROC curves in Figure \ref{fig:roc_multispecies}, and the PR curves in Figure \ref{fig:pr_multispecies}. All of the curves are plotted as a result of derived metrics from the raw softmax outputs averaged over BNN samples in Figure \ref{fig:softmax_multispecies}. We note overall healthy model outputs, with the probability space well occupied on a scale from 0 to 1. We can also inspect that, generally, predictions are concentrated in the correct regions: the model output per class mimics the true underlying label well, across all classes. This results in healthy ROC curves for all the classes. However, due to the class imbalance encountered, and possibly the difficulty of distinguishing certain species, PR scores leave room for improvement, especially in classes 5, 6 and 7, where the most confusion between species occurs (see confusion matrix of Figure \ref{fig:cm_multispecies}). Confusion occurs between species of similar physical characteristics, resulting in similar flight tones. We illustrate a random sample of spectrograms created from audio clips for all species from the IHI Tanzanian cup data in Figure \ref{fig:spec_multispecies}. We also supply a \href{https://github.com/HumBug-Mosquito/HumBugDB/blob/master/notebooks/spec_audio_multispecies.ipynb}{Jupyter tool} to visualise and to listen to mosquito samples of any species conveniently (illustrated in Figure \ref{fig:datavis_multispecies}).

It should be noted that as the mosquitoes are all wild individuals, it is natural that the variation within their species produces some difficulty for the models. Nevertheless, the ROC curves demonstrate that choosing model thresholds in combination with uncertainty estimation from the BNN arm us with the ability to perform species classification.
\begin{figure}
    \centering
 \begin{subfigure}[t]{1.0\textwidth}
\centering
      \includegraphics[width=0.8\textwidth]{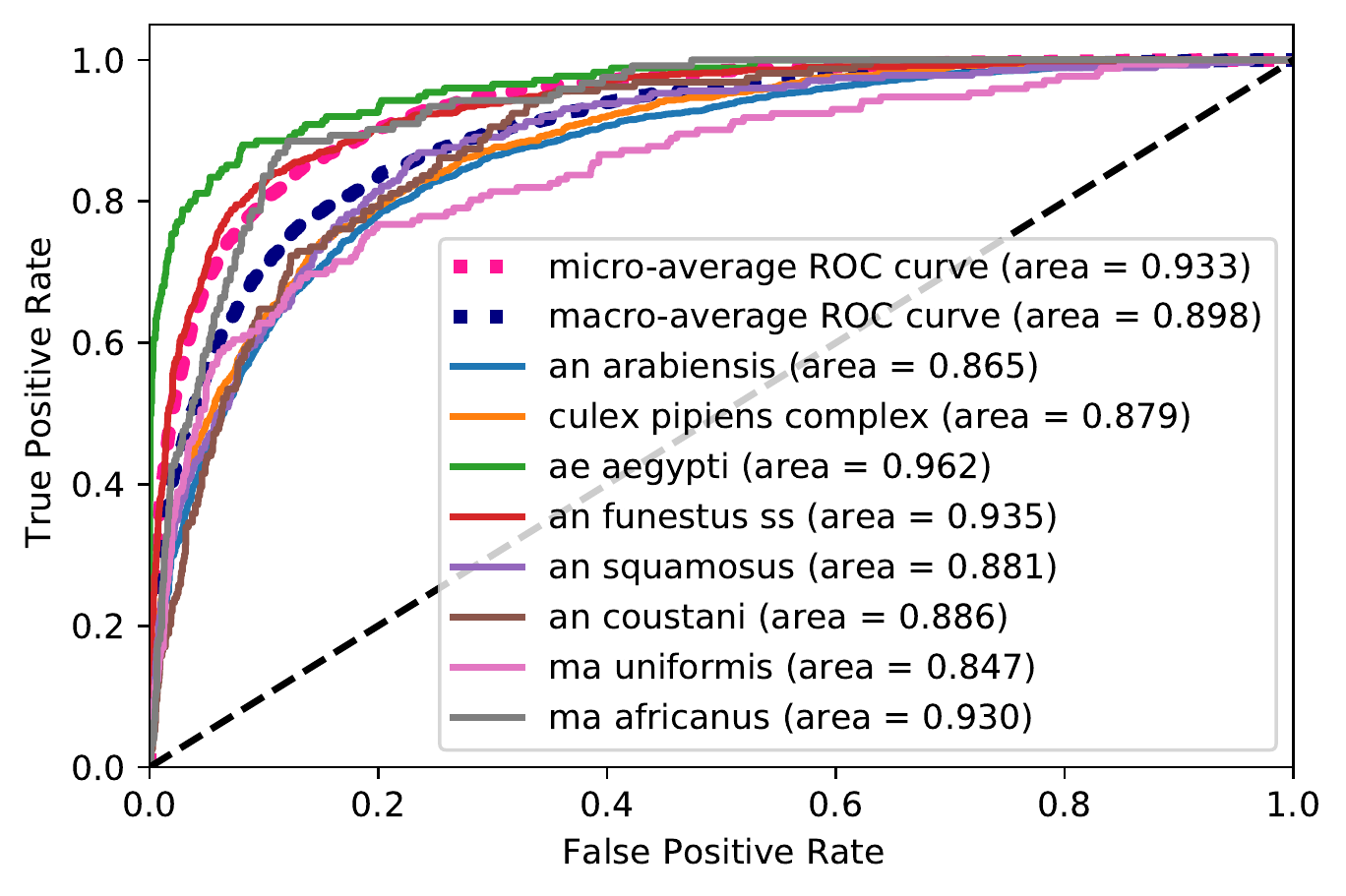}
    \caption{ROC curves and areas.}
    \label{fig:roc_multispecies}
    \end{subfigure}
    
    \begin{subfigure}[t]{1.0\textwidth}
    \centering
    \includegraphics[trim= 0 0 0 20, clip,width=1.0\textwidth]{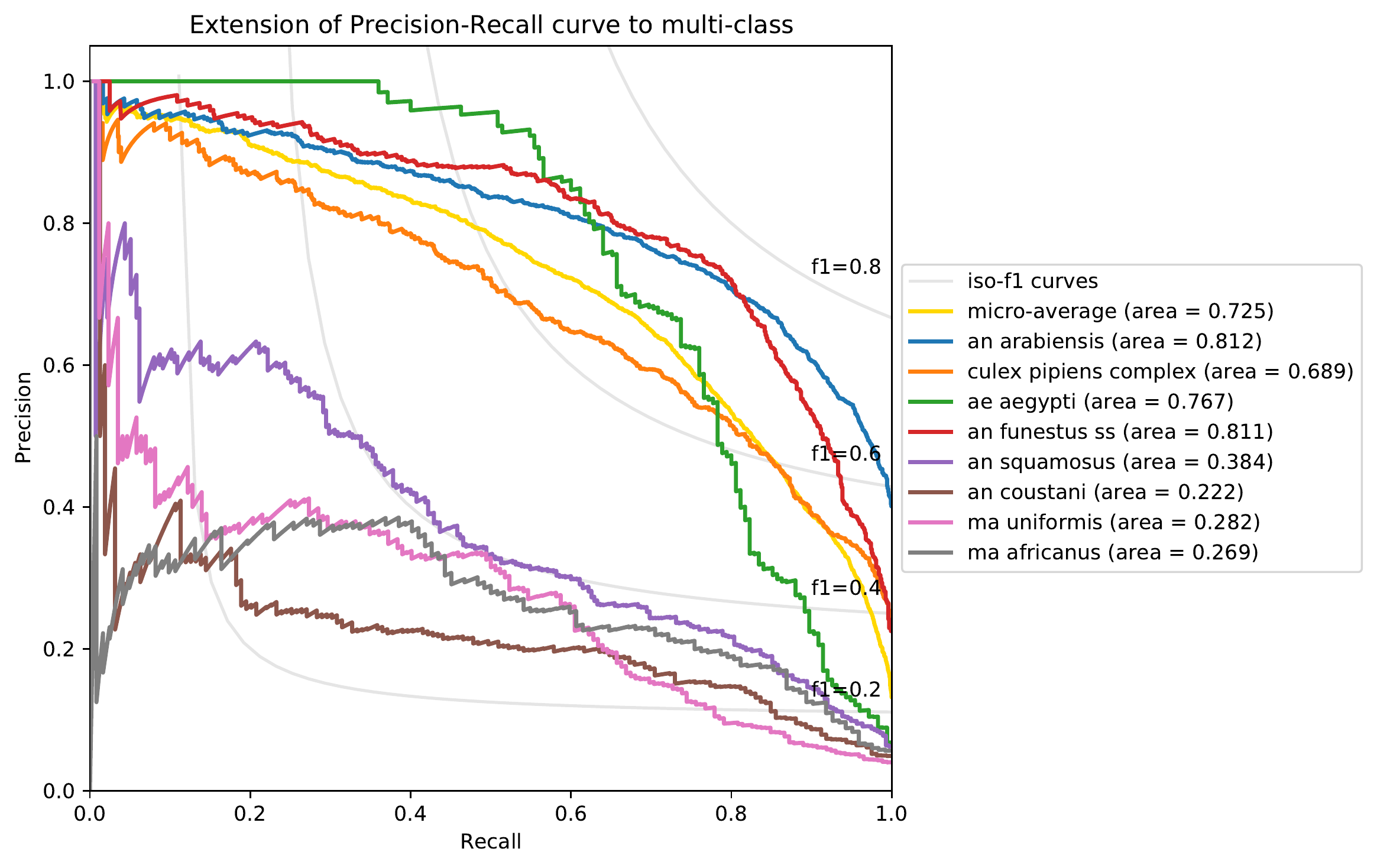}
    \caption{Precision-Recall curves, alongside isometric $F1$ score curves. Note that a measure of quality of modelling is given by the PR-AUC areas supplied in the legend. A `good' area is a score that is significantly above the prevalence of the class. For a list of prevalences consult Table \ref{tab:multispecies}.}
    \label{fig:pr_multispecies}
      \end{subfigure}
     \caption{Receiver Operating Characteristic and Precision Recall curves of multi-species performance on IHI Tanzanian cup data of wild mosquito data. Results generated with random seed of 42, with MozzBNNv2 Feat. B. Corresponding confusion matrices and the raw softmax outputs per class are given in Figures \ref{fig:cm_multispecies} and \ref{fig:softmax_multispecies} respectively.} 
    \label{fig:cm_roc_multispecies}
\end{figure}

\begin{figure}
    \centering
    \includegraphics[width=1.0\textwidth]{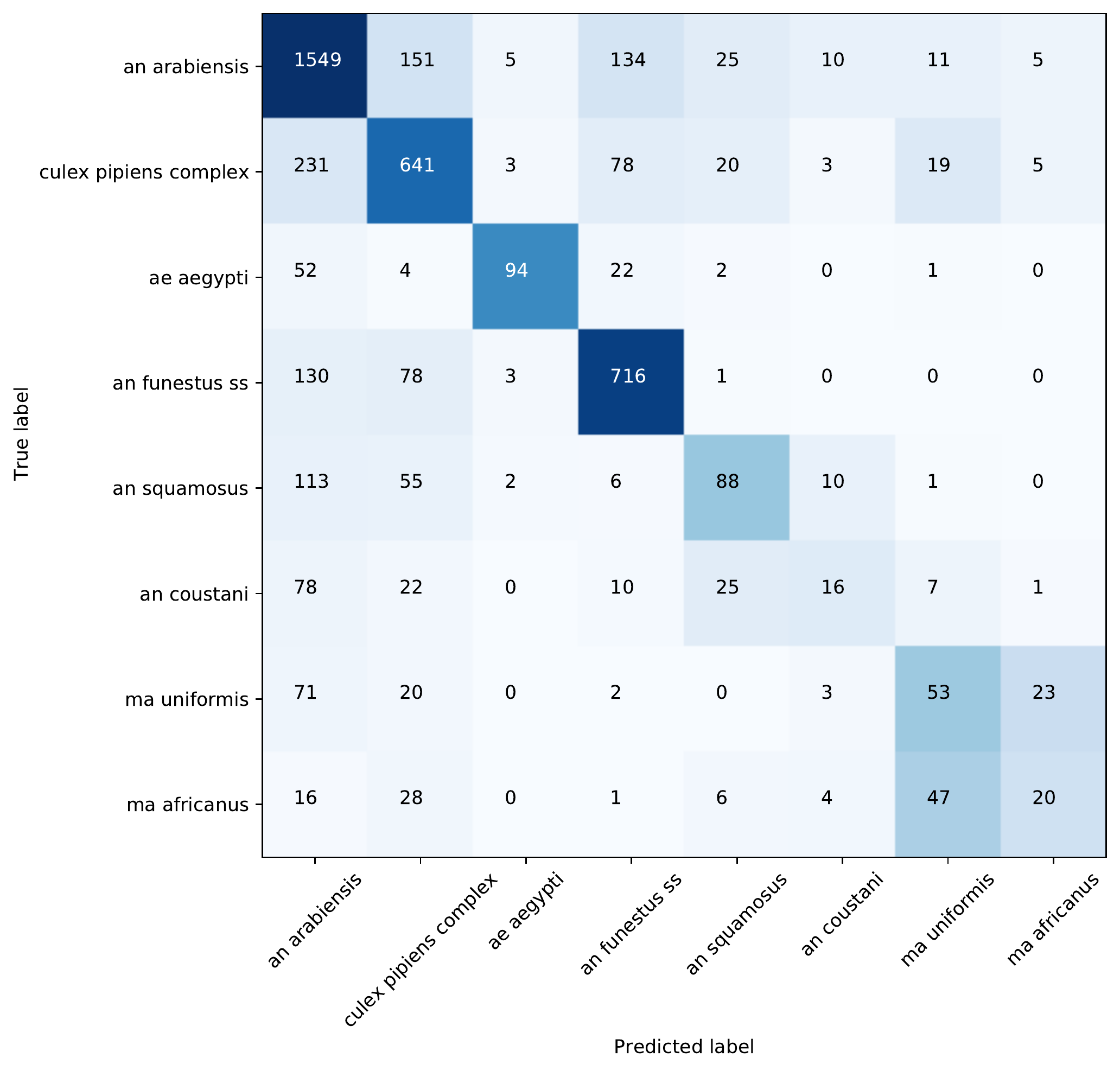}
    \caption{ Confusion matrix per species of IHI Tanzanian cup data of wild mosquito data. Results generated with random seed of 42, with MozzBNNv2 Feat. B. Each entry corresponds to a 1.92 second data sample, created with no overlap during feature extraction for the test data. The species of this matrix were arranged in neighbouring classes of similarity to more clearly illustrate where the confusion occurs. Confusion occurs between species of similar physical characteristics.}
    \label{fig:cm_multispecies}
\end{figure}

\begin{figure}
    \centering
    \includegraphics[width=1.0\textwidth]{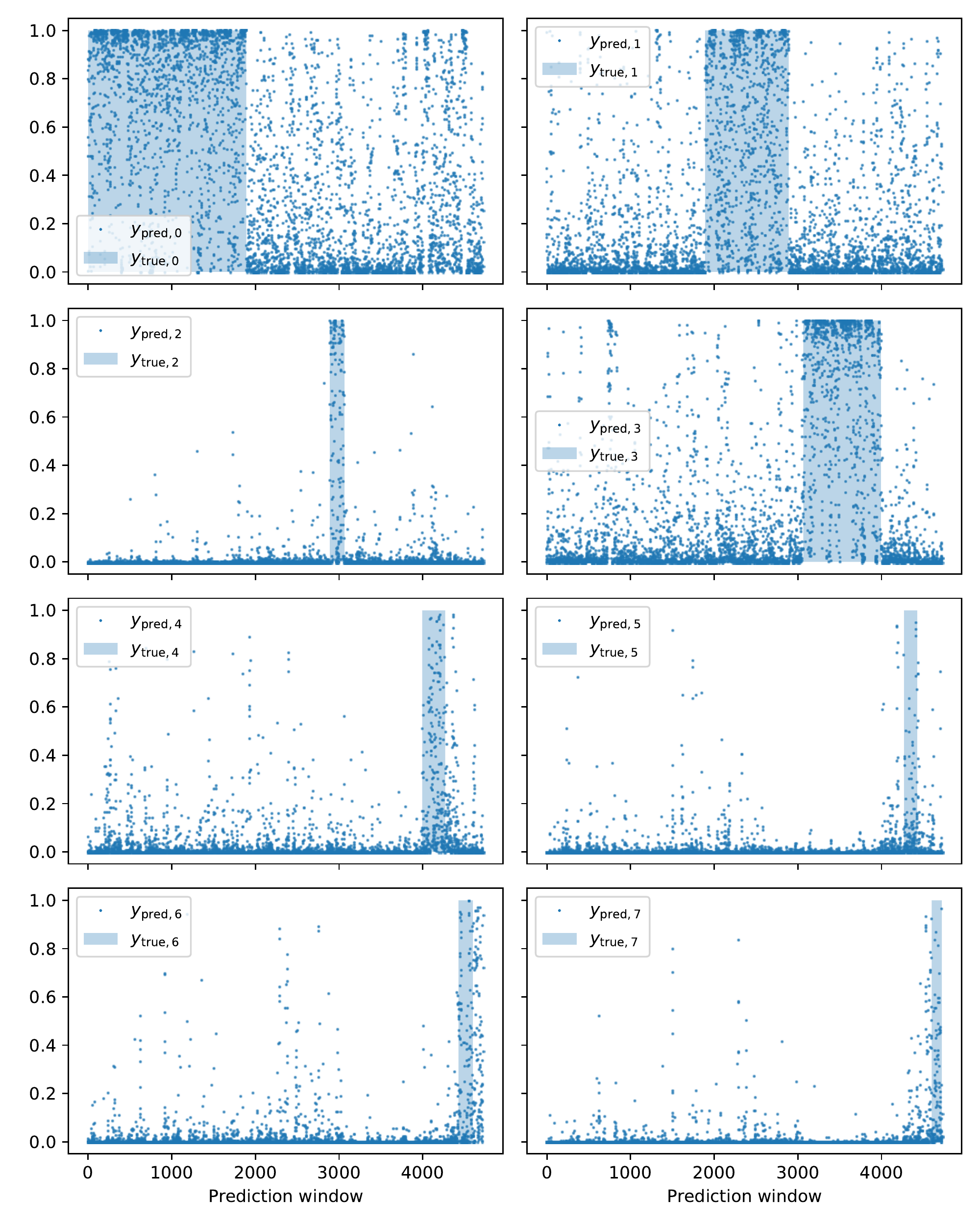}
    \caption{Raw softmax output, averaged across MC dropout samples, per species of IHI Tanzanian cup data of wild mosquito data. The shaded region represents the correct feature windows per class, and the dots are the algorithm predictions. Each plot represents a single class output of the final softmax layer. Results generated with random seed of 42, with MozzBNNv2 Feat. B. Species are numbered in order of classes of the confusion matrix.}
    \label{fig:softmax_multispecies}
\end{figure}

\begin{figure}
    \centering
    \includegraphics[width=1.0\textwidth]{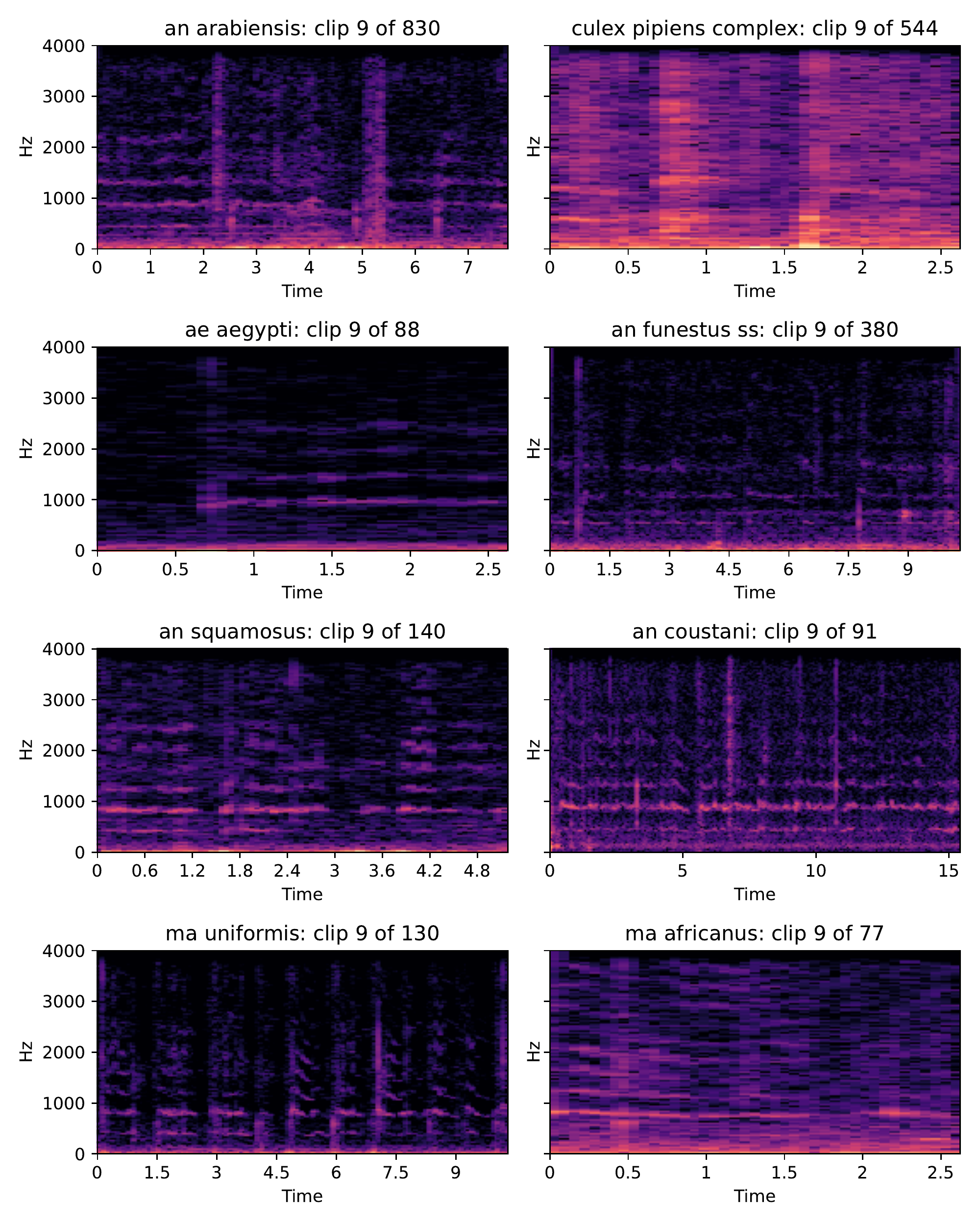}
    \caption{A comparison of randomly chosen clips of mosquito species of the IHI Tanzanian cup recordings. As seen from Figure \ref{fig:cm_multispecies}, model confusion occurs commonly between \textit{ma. uniformis} and \textit{ma. africanus} classes. On the other hand, \textit{ae. aegypti} are most easily identified, despite their relatively infrequent occurence in the dataset. Spectrograms constructed by re-sampling audio to 8\,kHz, matching the sample rate of Feat B. and the native sample rate of the smartphone app. In this format, it is also easier to visualise differences between species. For higher sample rate visualisations and audio, consult Figure \ref{fig:datavis_multispecies}.}
    \label{fig:spec_multispecies}
\end{figure}

\begin{figure}
    \centering
    \includegraphics[width=1.0\textwidth]{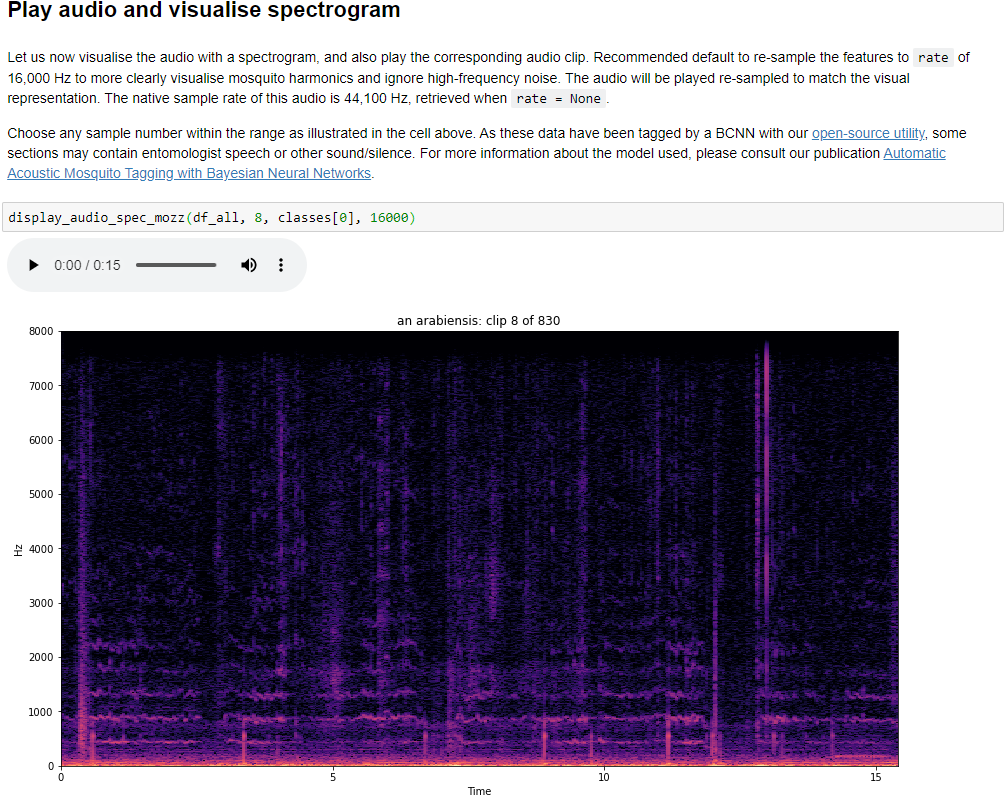}
    \caption{Interactive data visualisation tool on \url{https://github.com/HumBug-Mosquito/HumBugDB/blob/master/notebooks/spec_audio_multispecies.ipynb}. The user supplies a sample number to index an audio clip of any mosquito species from a \texttt{pandas} DataFrame. A spectrogram is displayed, alongside an audio playback component.}
    \label{fig:datavis_multispecies}
\end{figure}

\clearpage
\section{PostgreSQL Database}
\label{sec:appendix_db_metadata}
\subsection{Database metadata}
The data presented in this paper are regularly maintained in a PostgreSQL database. For completeness, we include the full schema in Figure \ref{fig:sql}. We note that since data upload is a constant work in progress, some fields have not yet been populated sufficiently to be useful upon data extraction. We thus restrict the metadata to the fields that have been verified, and are most likely  to be of greatest use. The command we use to extract all the metadata for this paper is as follows:

\begin{lstlisting}[
           language=SQL,
           showspaces=false,
           breaklines = true,
           numbers=left,
           numberstyle=\tiny,
           morekeywords={\copy},
           commentstyle=\color{gray}
        ]
\copy (SELECT label.id, fine_end_time-fine_start_time, name, sample_rate, record_datetime, sound_type, species, gender, fed, plurality, age, method, mic_type, device_type, country, district, province, place, location_type
FROM label
LEFT JOIN mosquito ON (label.mosquito_id = mosquito.id)
RIGHT JOIN audio ON (label.audio_id = audio.id)
RIGHT JOIN device ON (audio.dev_id = device.id)
WHERE type = 'Fine' 
AND fine_start_time IS NOT NULL AND sound_type in
('mosquito', 'background', 'audio', 'wasp','fly') AND
(path LIKE '%Kenya%'
OR path LIKE '%Thai%'
OR path LIKE '%Tanzania%'
OR path LIKE '%LSTMH%'
OR path LIKE '%CDC%'
OR path LIKE '%Culex%')
ORDER BY path) to '/data/export/neurips_2021_zenodo_0_0_1.csv' csv header;
\end{lstlisting}

We will now break down each metadata field in the data release by the table it originated from and its column heading.

\textbf{Label:}
\begin{itemize}
    \item \texttt{label.id} selects the column \texttt{id} from the table \texttt{label}, which is joined to \texttt{audio} on \texttt{label.audio\_id = audio.id}. This allows us to now extract a labelled section of audio as indicated by the start and end times of the label. 
    \item \texttt{fine\_start\_time}, \texttt{fine\_end\_time} are the tags for start and end of the audio label, with reference to the original audio recording. Once audio is extracted, we assign the labelled section the filename set to the \texttt{label.id}, and define a column \texttt{length} which takes the value of \texttt{fine\_start\_time - fine\_end\_time} for each new label.
\end{itemize}
\textbf{Audio:}
\begin{itemize}
    \item \texttt{name}: The original filename of the recording (including file extension).
  \item \texttt{sample\_rate}: The sample rate of the recording.
\item \texttt{record\_datetime} The time of recording, as SQL \texttt{DATETIME} object (easy to parse with either \texttt{pandas} or built-in \texttt{datetime.datetime}). For newer data, this timestamp is exact, however data collected prior may only be correct to the month.
\end{itemize}
\textbf{Mosquito:}
\begin{itemize}
    \item \texttt{species} is the species of the mosquito, either the species complex, or more specifically the species if available (e.g. \textit{An. arabiensis} of the complex \textit{An. gambiae s.s.}). If no species information is available, this field is blank (or \texttt{NaN} when imported by \texttt{pandas} with default settings). A full breakdown of the available species per experimental group is given in Figure \ref{fig:species_dist} and Table \ref{tab:appendix_species}.    
        \item \texttt{gender}: Gender of mosquitoes (\texttt{M} or \texttt{F}) or blank if not known.
    \item \texttt{fed}: Whether mosquito has been fed (\texttt{t} or \texttt{f}) or blank.
    \item \texttt{plurality}: The quantity of mosquitoes recorded at one instant: \texttt{single}, \texttt{plural} or blank if unknown.
    \item \texttt{age}: The age of mosquito in days.
    \item \texttt{sound\_type}: denotes whether the label corresponds to a mosquito event if \texttt{mosquito}, but can take the value of \texttt{background} for corresponding background, \texttt{audio} for sections of dense audio events not containing mosquito or \texttt{wasp} and \texttt{fly}. When parsing data, a binary distinction between \texttt{mosquito} and  \texttt{NOT mosquito} can be made safely.
\end{itemize}
\textbf{Device:}
\begin{itemize}
    \item\texttt{method}: The method of capture of mosquitoes, taking values \texttt{HBN}, \texttt{LT}, \texttt{ABN}, \texttt{LC}, \texttt{HLC} or none if not known (or applicable).   
    Human-baited nets (\texttt{HBN}) are a form of mosquito intervention where humans are surrounded by a mosquito net. As part of the HumBug project, adapted bednets were used where an additional canopy to hold smartphones for recording was sewn on (from 2020 onwards) \citep{2021MEE}.
    
    Animal-baited nets follow the same concept but involve an animal as the main attractant for mosquitoes.
    
    CDC light traps (\texttt{LT}) use several attractants to lure mosquitoes into the collection chamber. Light is the primary source, but bottled CO\textsubscript{2}, gas or dry ice can also be used. 
    
    Larval collections, where the eggs of young mosquitoes are collected, are denoted \texttt{LC}.
    
    Human landing catches, where mosquitoes that landed on humans are caught, are denoted \texttt{HLC}.
    
    For mosquitoes raised from culture and not released into the wild and/or near any nets, this field is blank.
    \item\texttt{mic\_type}: The microphone used. Takes values \texttt{telinga}, \texttt{phone} to denote the microphone type. Use this field to filter audio by the type of sound produced, if you wish to check for bias arising from recording device. Further refine the search with the phone model as specified in \texttt{device\_type}.
    \item\texttt{device\_type}: the device to which the microphone was connected. E.g. the field microphone (Telinga) was connected to a \texttt{Tascam} or \texttt{Olympus} recorder. If a smartphone was used, the device is the phone model (e.g. \texttt{itel A16} or \texttt{Alcatel 4015X}). 
\end{itemize}
\textbf{Location:}
\begin{itemize}
    \item\texttt{location\_type}: The environment in which the mosquitoes were recorded in, taking values \texttt{cup} for mosquitoes recorded in sample cups, \texttt{field} for mosquitoes recorded free-flying in the field (applicable to Tanzania 2020 bednet recordings), or \texttt{culture} for mosquitoes recorded in culture cages.
    \item\texttt{country}, \texttt{district}, \texttt{province}, \texttt{place}: The country, district, province, and name of the recording site (e.g. \texttt{USA}, \texttt{Georgia}, \texttt{Atlanta}, \texttt{CDC insect culture, Atlanta}). Use these values combined with \texttt{location\_type} to filter data by recording experiment. 
\end{itemize}

\begin{figure}
    \centering
    \includegraphics[page=2,trim= 85 460 85 450, clip, width=1.0\textwidth]{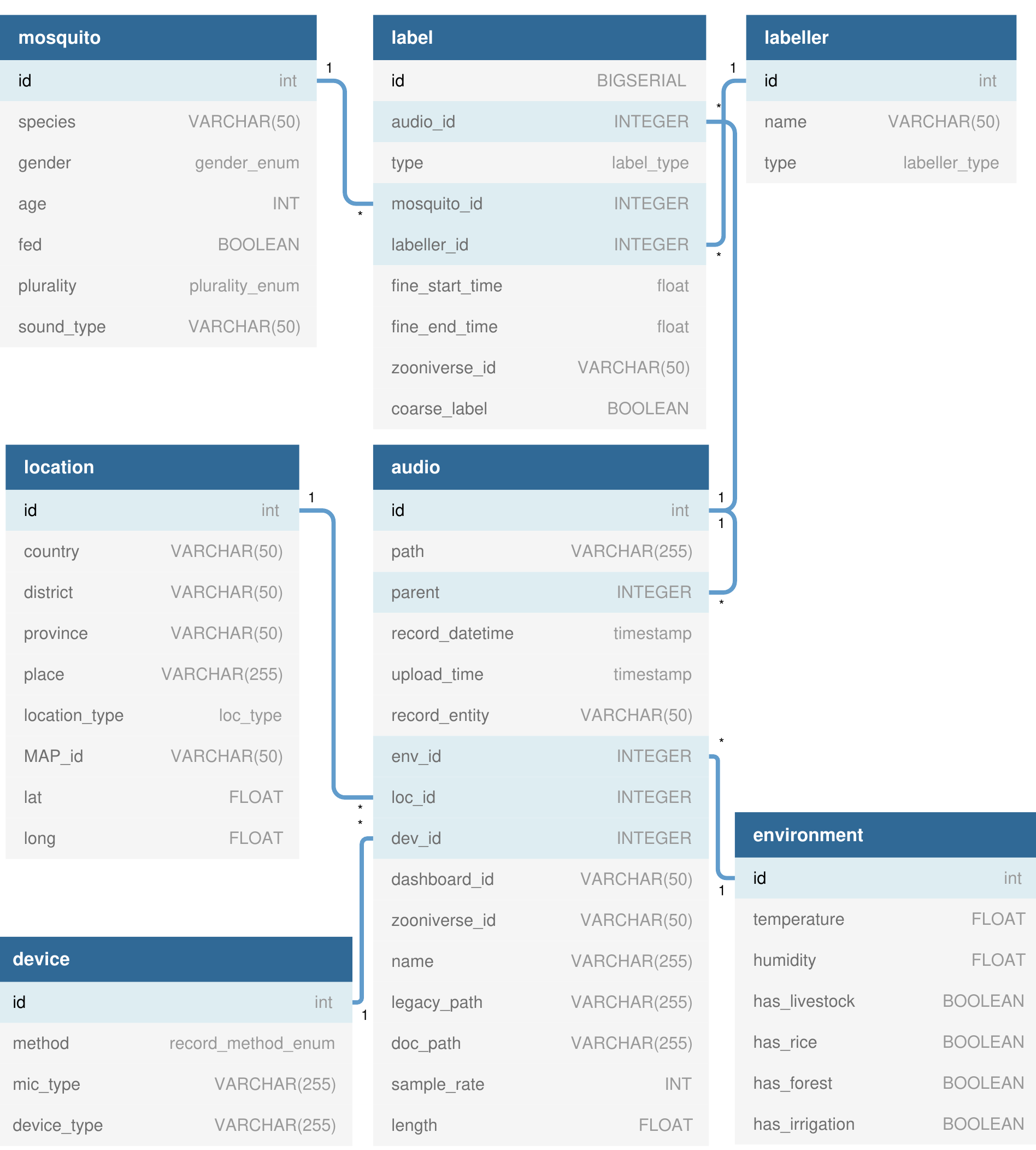}
    \caption{Relational tables of the full PostgreSQL database which was used to generate the data for this paper. The structured nature of the database enforces a standard in label format, ensuring we can efficiently mix and match data from a wide range of experiments with differing protocols. For example, if we wish to investigate the effect of mosquito gender or microphone type on the ability to detect mosquitoes, we may sub-select data with the appropriate metadata with one query. Database schema generated with \url{dbdiagram.io} from with \texttt{pg\_dump -s}.}
    \label{fig:sql}
\end{figure}

\begin{figure}
    \centering
    \includegraphics[width=1.\textwidth]{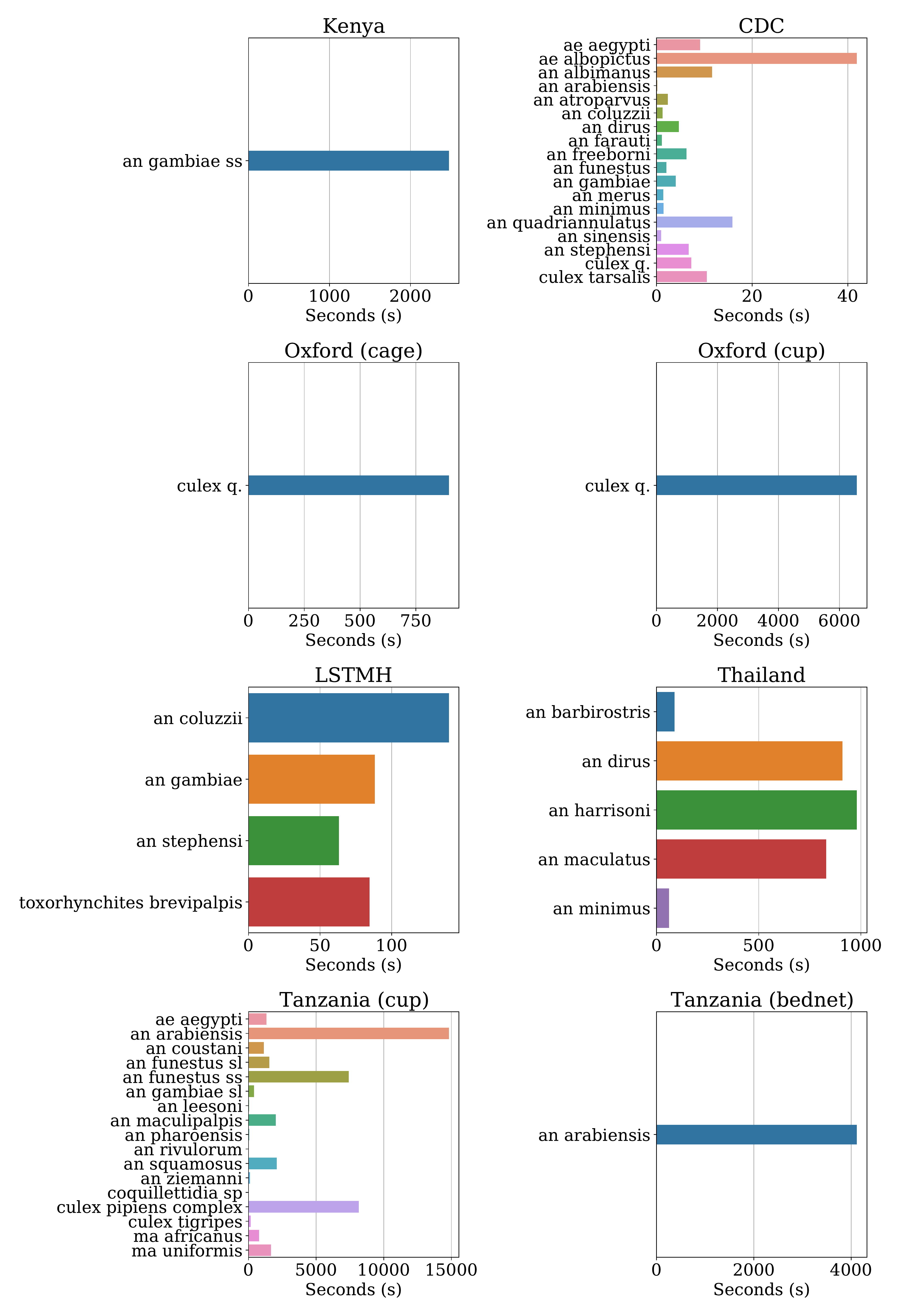}
    \caption{Species distribution per experiment corresponding to Table \ref{tab:Summary_for_appendix}.}
    \label{fig:species_dist}
\end{figure}

\newcolumntype{G}{>{\columncolor{Gray}}l}
\newcolumntype{R}{>{\columncolor{Gray}}r}
\begin{table}[]
\caption{Species distribution per experimental group corresponding to Table \ref{tab:Summary_for_appendix} and Figure \ref{fig:species_dist}.}
\label{tab:appendix_species}
\centering
\small
\begin{tabular}{@{}GRRRRRRRRR@{}}
\toprule \rowcolor{white}
\multirow{1}{*}{Species} & \multicolumn{9}{c}{Species per location (seconds)}                                                                                                                                                                                                                                                                                                                                                                                                                                                                        \\ \cmidrule(l){2-10} \rowcolor{white}
                         & \multicolumn{1}{r}{Kenya} & \multicolumn{1}{r}{\begin{tabular}[r]{@{}r@{}}USA\\ (CDC)\end{tabular}} & \multicolumn{1}{r}{\begin{tabular}[r]{@{}r@{}}Oxford\\ (cup)\end{tabular}} & \multicolumn{1}{r}{\begin{tabular}[r]{@{}r@{}}Oxford \\ (cage)\end{tabular}} & \multicolumn{1}{r}{LSTMH} & \multicolumn{1}{r}{Thailand} & \multicolumn{1}{r}{\begin{tabular}[r]{@{}r@{}}Tanzania \\ (cup)\end{tabular}} & \multicolumn{1}{r}{\begin{tabular}[r]{@{}r@{}}Tanzania\\  (bednet)\end{tabular}} & \textbf{Total} \\ \midrule
\textit{Ae. aegypti }               & 0                         & 9.1                                                                       & 0                                                                          & 0                                                                            & 0                         & 0                          & 1322.4                                                                             & 0                                                                                & \textbf{1333.6}                        \\
\rowcolor{white} \textit{Ae. albopictus  }           & 0                         & 41.9                                                                       & 0                                                                          & 0                                                                            & 0                         & 0                            & 0                                                                             & 0                                                                                & \textbf{41.9}                         \\
\textit{An. albimanus  }          & 0                         & 11.6                                                                       & 0                                                                          & 0                                                                            & 0                         & 0                            & 0                                                                             & 0                                                                                & \textbf{11.6}                       \\
\rowcolor{white} \textit{An. arabiensis }             & 0                         & 0.1                                                                       & 0                                                                          & 0                                                                            & 0                         & 0                            & 14815.2                                                                             & 4118.2                                                                                & \textbf{18933.6}                         \\
\textit{An. atroparvus   }             & 0                         & 2.3                                                                       & 0                                                                          & 0                                                                            & 0                         & 0                            & 0                                                                             & 0                                                                                & \textbf{2.4}                       \\
\rowcolor{white} \textit{An. barbirostris }              & 0                         & 0                                                                       & 0                                                                          & 0                                                                            & 0                         & 87.8                            & 0                                                                             & 0                                                                                & \textbf{87.8}                      \\
\textit{An. coluzzii  }            & 0                         & 1.3                                                                       & 0                                                                          & 0                                                                            & 140.0                         & 0                            & 0                                                                             & 0                                                                                & \textbf{141.3}                        \\
\rowcolor{white} \textit{An. coustani     }           & 0                         & 0                                                                       & 0                                                                          & 0                                                                            & 0                         & 0                            & 1140.6                                                                             & 0                                                                                & \textbf{1140.6}                        \\
\textit{An. dirus}               & 0                         & 4.7                                                                       & 0                                                                          & 0                                                                            & 0                         & 909.8                          & 0                                                                             & 0                                                                                & \textbf{914.5}                        \\
\rowcolor{white} \textit{An. farauti }              & 0                         & 1.1                                                                       & 0                                                                          & 0                                                                            & 0                         & 0                            & 0                                                                             & 0                                                                                & \textbf{1.1}                        \\
\textit{An. freeborni  }             & 0                         & 6.3                                                                       & 0                                                                          & 0                                                                            & 0                         & 0                            & 0                                                                             & 0                                                                                & \textbf{6.2}                         \\
\rowcolor{white} \textit{An. funestus }           & 0                         & 2.1                                                                       & 0                                                                          & 0                                                                            & 0                         & 0                            & 0                                                                             & 0                                                                                & \textbf{2.1}                        \\
\textit{An. funestus s.l. }              & 0                         & 0                                                                       & 0                                                                          & 0                                                                            & 0                         & 0                            & 1542.1                                                                             & 0                                                                                & \textbf{1542.1}                         \\
\rowcolor{white} \textit{An. funestus s.s. }              & 0                         & 0                                                                       & 0                                                                          & 0                                                                            & 0                         & 0                            & 7414.2                                                                             & 0                                                                                & \textbf{7414.2}                         \\
\textit{An. gambiae   }          & 0                         & 4.0                                                                      & 0                                                                          & 0                                                                            & 88.2                         & 0                            & 0                                                                             & 0                                                                                & \textbf{92.2}                        \\
\rowcolor{white} \textit{An. gambiae s.l.  }          & 0                         & 0                                                                       & 0                                                                          & 0                                                                            & 0                         & 0                            & 406.7                                                                             & 0                                                                                & \textbf{406.7}                         \\
\textit{An. gambiae s.s.   }            & 2474.2                         & 0                                                                       & 0                                                                          & 0                                                                            & 0                         & 0                            & 0                                                                             & 0                                                                                & \textbf{2474.2}                         \\
\rowcolor{white}\textit{An. harrisoni }              & 0                         & 0                                                                       & 0                                                                          & 0                                                                            & 0                         & 980.4                           & 0                                                                             & 0                                                                                & \textbf{980.4}                         \\
\textit{An. leesoni    }         & 0                         & 0                                                                       & 0                                                                          & 0                                                                            & 0                         & 0                            & 43.5                                                                            & 0                                                                                & \textbf{43.5}                        \\
\rowcolor{white}\textit{An. maculatus   }           & 0                         & 0                                                                       & 0                                                                          & 0                                                                            & 0                         & 830.4                           & 0                                                                             & 0                                                                                & \textbf{830.4}                         \\
\textit{An. maculipalpis }             & 0                         & 0                                                                       & 0                                                                          & 0                                                                            & 0                         & 0                            & 2013.0                                                                             & 0                                                                                & \textbf{2013.0}                         \\
\rowcolor{white} \textit{An. merus   }        & 0                         & 1.4                                                                     & 0                                                                          & 0                                                                            & 0                         & 0                            & 0                                                                             & 0                                                                                & \textbf{1.4}                         \\
\textit{An. minimus }              & 0                         & 1.4                                                                   & 0                                                                          & 0                                                                            & 0                         & 61.5                          & 0                                                                             & 0                                                                                & \textbf{63.0}                        \\
\rowcolor{white} \textit{An. pharoensis  }            & 0                         & 0                                                                       & 0                                                                          & 0                                                                            & 0                         & 0                            & 56.3                                                                             & 0                                                                                & \textbf{56.3}                         \\
\textit{An. quadriannulatus }             & 0                         & 15.9                                                                       & 0                                                                          & 0                                                                            & 0                         & 0                            & 0                                                                             & 0                                                                                & \textbf{15.9}                         \\
\rowcolor{white} \textit{An. rivulorum  }   & 0                         & 0                                                                       & 0                                                                          & 0                                                                            & 0                         & 0                            & 5.1                                                                             & 0                                                                                & \textbf{5.1}                         
\\
\textit{An. sinensis  }             & 0                         & 1.0                                                                       & 0                                                                          & 0                                                                            & 0                         & 0                            & 0                                                                             & 0                                                                                & \textbf{1.0}        
\\
\rowcolor{white} \textit{An. squamosus }             & 0                         & 0                                                                       & 0                                                                          & 0                                                                            & 0                         & 0                            & 2091.8                                                                             & 0                                                                                & \textbf{2091.8}        
\\
\textit{An. stephensi  }           & 0                         & 6.7                                                                     & 0                                                                          & 0                                                                            & 63.1                       & 0                            & 0                                                                             & 0                                                                                &  \textbf{69.9}       
\\
\rowcolor{white} \textit{An. ziemanni}               & 0                         & 0                                                                       & 0                                                                          & 0                                                                            & 0                         & 0                            & 110.0                                                                             & 0                                                                                & \textbf{110.0}        
\\
\textit{Coquillettidia sp.}             & 0                         & 0                                                                       & 0                                                                          & 0                                                                            & 0                         & 0                            & 25.6                                                                           & 0                                                                                & \textbf{25.6}       
\\
\rowcolor{white} \textit{Culex pipiens }            & 0                         & 0                                                                       & 0                                                                          & 0                                                                            & 0                         & 0                            & 8157.8                                                                              & 0                                                                                & \textbf{8157.8}
\\
\textit{Culex q.  }          & 0                         & 7.3                                                                       & 6573.1 &      898.1                                                                                                                                          & 0                         &             0              & 0                                                                             & 0                                                                                & \textbf{7478.5}        
\\
\rowcolor{white}  \textit{Culex tarsalis }           & 0                         & 10.5                                                                       & 0                                                                          & 0                                                                            & 0                         & 0                            & 0                                                                             & 0                                                                                & \textbf{10.5}        
\\
\textit{Culex tigripes }          & 0                         & 0                                                                       & 0                                                                          & 0                                                                            & 0                         & 0                            & 158.7                                                                             & 0                                                                                & \textbf{158.7}        
\\
\rowcolor{white} \textit{Ma. africanus }         & 0                         & 0                                                                       & 0                                                                          & 0                                                                          & 0                         & 0                            & 785.2                                                                               & 0                                                                                & \textbf{785.2}        
\\
\textit{Ma. uniformis }           & 0                         & 0                                                                       & 0                                                                          & 0                                                                            & 0                         & 0                            & 1654.5                                                                           & 0                                                                                & \textbf{1654.6}        
\\
\rowcolor{white}  \textit{Toxorhynchites br.}          & 0                         & 0                                                                       & 0                                                                          & 0                                                                            & 84.6                         & 0                            & 0                                                                             & 0                                                                                & \textbf{84.6}        
\\
\bottomrule
\end{tabular}
\end{table}

\clearpage
\subsection{Miscellaneous commands}

To generate the metadata of Table \ref{tab:Summary_for_appendix}, we include a list of commands used to generate one row for completeness.

Count total length of labelled audio for a certain path and sound type:
\begin{lstlisting}[
           language=SQL,
           showspaces=false,
           breaklines = true,
           numbers=left,
           numberstyle=\tiny,
           morekeywords={\copy},
           commentstyle=\color{gray}
        ]
SELECT SUM(fine_end_time-fine_start_time) 
FROM label 
LEFT JOIN mosquito ON (label.mosquito_id = mosquito.id)
RIGHT JOIN audio ON (label.audio_id = audio.id)
RIGHT JOIN location ON (audio.loc_id = location.id) 
WHERE path LIKE '%Thai%' and sound_type='mosquito';
\end{lstlisting}

Count number of audio files for a certain path and sound type:
\begin{lstlisting}[
           language=SQL,
           showspaces=false,
           breaklines = true,
           numbers=left,
           numberstyle=\tiny,
           morekeywords={\copy},
           commentstyle=\color{gray}
        ]
SELECT COUNT (DISTINCT path) 
FROM label 
LEFT JOIN mosquito ON (label.mosquito_id = mosquito.id) 
RIGHT JOIN audio ON (label.audio_id = audio.id) 
RIGHT JOIN location ON (audio.loc_id = location.id)
WHERE path LIKE '%Thai%' and sound_type='mosquito';
\end{lstlisting}
Return location, device types, and recording methods for dataset:
\begin{lstlisting}[
           language=SQL,
           showspaces=false,
           breaklines = true,
           numbers=left,
           numberstyle=\tiny,
           morekeywords={\copy},
           commentstyle=\color{gray}
        ]
SELECT DISTINCT country, location_type, method, mic_type, device_type
FROM audio
RIGHT JOIN location ON (audio.loc_id = location.id)
RIGHT JOIN device ON (audio.dev_id = device.id)
WHERE path LIKE '%Tanzania%';
\end{lstlisting}

\section{Datasheet for HumBugDB}
\label{sec:appendix_datasheet_for_dataset}
We follow the structure outlined in Datasheets for Datasets by \citet{gebru2018datasheets}. \ref{sec:appendix_motivation} gives the motivation for the data. \ref{sec:appendix_composition} describes the composition of the data. \ref{sec:appendix_datasheet_collection} describes the collection process. \ref{sec:appendix_datasheet_preprocessing} describes the preprocessing involved in the data curation. \ref{sec:appendix_datasheet_uses} lists past uses, and suggests a range of future uses in depth. \ref{sec:appendix_datasheet_databias} describes potential sources of data bias and relevant mitigation strategies. Database maintenance policies are given in \ref{sec:appendix_database_maintenance}.


\subsection{Motivation}
\label{sec:appendix_motivation}
\paragraph{For what purpose was the dataset created?}
This dataset was created for academic research, and applications of machine learning for global health. One such application is the monitoring of deadly mosquito species from their acoustic signature, for which quality training data is required to capture the variation that species may exhibit.
\paragraph{Who created the dataset (e.g., which team, research group) and on
behalf of which entity (e.g., company, institution, organization)?}
This dataset was curated by the Machine Learning Research Group of the University of Oxford. Data was collected by the Department of Zoology, University of Oxford, the Centers for Disease Control and Prevention, Atlanta, the United States Army Medical Research Unit in Kenya (USAMRU-K), at the London School of Tropical Medicine and Hygiene, the Dept of Entomology, Kasesart University, Bangkok, and by the Ifakara Health Institute in Tanzania.
\paragraph{Who funded the creation of the dataset?}
A Google Impact Challenge Award 2014, The Bill and Melinda Gates Foundation (2019--present), available on \url{https://www.gatesfoundation.org/about/committed-grants/2019/07/opp1209888} (last accessed: June 2021).
\subsection{Composition}
\label{sec:appendix_composition}
\paragraph{What do the instances that comprise the dataset represent (e.g.,
documents, photos, people, countries)?} This dataset is a collection of acoustic recordings in \texttt{wav} PCM format. We also supply all the metadata, generated in PostgreSQL to a \texttt{csv} file.

\paragraph{How many instances are there in total (of each type, if appropriate)?}
9,295 \texttt{wav} audio files, 1 \texttt{csv}.

\paragraph{Does the dataset contain all possible instances or is it a sample
(not necessarily random) of instances from a larger set?} The audio files are a sub-sample of complete audio recordings, with the recordings corresponding to one complete label defined with a label ID, extracted from the original audio with the markers \texttt{start\_time, end\_time}. We are unable to release the full unlabelled audio due to potential issues with privacy and personally identifiable information. The metadata is a curated sub-sample of all available metadata, where fields which were not sufficiently populated or unverified are excluded.

\paragraph{What data does each instance consist of?}
Each instance corresponds to a labelled section of audio with the event times originally tagged in the original recording with a \texttt{start\_time, end\_time}, either manually by human domain experts, or by machine learning models. The label type is supplied in the metadata.

\paragraph{Is there a label or target associated with each instance?}

Yes, every recording matches a label.
\paragraph{Is any information missing from individual instances? }

Though every single sample has a label, some recordings have greater availability of metadata than others; see the metadata \texttt{csv} for details.
\FloatBarrier
\paragraph{Are there recommended data splits (e.g., training, development/validation,
testing)? } 
Yes, see Table \ref{tab:Summary_for_appendix}. The splits are carried out to increase the chance of generalisation to recordings conducted in varying conditions. The validation split is part of the challenge of this benchmark, left to the discretion of the users. The test data is automatically split in the supplied code.
The two tasks that the data splits encouraged are defined as follows:
\begin{itemize}
\item Mosquito Event Detection (MED): distinguishing mosquitoes of any species from their background surroundings, such as other insects, speech, urban, and rural noise. 
\item Mosquito Species Classification (MSC):  the classification of detected mosquitoes into their respective species.
\end{itemize}

\begin{table}
\centering
\footnotesize
\caption{Key audio metadata and division into train/test for the tasks of MED: Mosquito Event Detection, and MSC: Mosquito Species Classification.  \textit{`Wild'} mosquitoes  captured and placed into paper \textit{`cups'} or attracted by bait surrounded by \textit{`bednets'}. \textit{`Culture'} mosquitoes bred specifically for research. Total length (in seconds) of mosquito recordings per group given, with the availability of species meta-information in parentheses. Total length of corresponding non-mosquito recordings, with matching environments, given as \textit{`Negative'}. Full metadata documented in Appendix \ref{sec:appendix_db_metadata}.}
\label{tab:Summary_for_appendix}
\begin{tabular}{@{}GGGGGGG@{}}
\toprule
\rowcolor{white} 
\begin{tabular}[l]{@{}l@{}}Tasks:\\ Train/Test\end{tabular} &\begin{tabular}[l]{@{}l@{}}Mosquito\\ origin\end{tabular}                    &  \begin{tabular}[l]{@{}l@{}}Site\\ Country\end{tabular}                                                        & \begin{tabular}[l]{@{}l@{}}Method\\ (year)\end{tabular} & \begin{tabular}[l]{@{}l@{}}Device\\ (sample rate)\end{tabular} & \begin{tabular}[c]{@{}c@{}}Mosquito (s) \\ (with species)\end{tabular} & \begin{tabular}[l]{@{}l@{}}Negative\\ (s)\end{tabular} \\ 
\midrule
 \begin{tabular}[l]{@{}l@{}}MSC: Train/Test \\ MED: Train\end{tabular}  &Wild                     & \begin{tabular}[l]{@{}l@{}}IHI \\ Tanzania\end{tabular}   & \begin{tabular}[l]{@{}l@{}}Cup \\ (2020)\end{tabular}         & \begin{tabular}[l]{@{}l@{}}Telinga\\ 44.1\,kHz\end{tabular}         & \begin{tabular}[l]{@{}l@{}}45,998\\ 45,998\end{tabular}                      &    5,600      \\  \midrule   

\rowcolor{white}
MED: Train  &Wild    & \begin{tabular}[l]{@{}l@{}}Kasetsart  \\ Thailand\end{tabular}  & \begin{tabular}[l]{@{}l@{}}Cup \\ (2018)\end{tabular}                                                & \begin{tabular}[l]{@{}l@{}}Telinga\\ 44.1\,kHz\end{tabular}           & \begin{tabular}[l]{@{}l@{}}9,306\\ 2,869\end{tabular}          & 7,896     \\  
 
MED: Train &Culture &   \begin{tabular}[l]{@{}l@{}}OxZoology\\ UK\end{tabular}       & \begin{tabular}[l]{@{}l@{}}Cup \\ (2017)\end{tabular}          &           \begin{tabular}[l]{@{}l@{}}Telinga\\ 44.1\,kHz\end{tabular}                                                     & \begin{tabular}[l]{@{}l@{}}6,573\\ 6,573\end{tabular}                   & 1,817     \\  

\rowcolor{white}

MED: Train &Culture & \begin{tabular}[l]{@{}l@{}}LSTMH \\  (UK)\end{tabular}         & \begin{tabular}[l]{@{}l@{}}Cup \\ (2018)\end{tabular}         &                              \begin{tabular}[l]{@{}l@{}}Telinga\\ 44.1\,kHz\end{tabular}                                   & \begin{tabular}[l]{@{}l@{}}376\\ 376\end{tabular}                    & 147      \\ 
              MED: Train&Culture       & \begin{tabular}[l]{@{}l@{}}CDC \\ USA\end{tabular}           & \begin{tabular}[l]{@{}l@{}}Cage \\ (2016)\end{tabular}        & \begin{tabular}[l]{@{}l@{}}Phone\\ 8\,kHz\end{tabular}                                                               & \begin{tabular}[l]{@{}l@{}}133\\ 127\end{tabular}         & 1,121     \\ 
                               \rowcolor{white}
              MED: Train &Culture        & \begin{tabular}[l]{@{}l@{}}USAMRU\\  Kenya\end{tabular}            & \begin{tabular}[l]{@{}l@{}}Cage \\ (2016)\end{tabular}     &                   \begin{tabular}[l]{@{}l@{}}Phone\\ 8\,kHz\end{tabular}                                               & \begin{tabular}[l]{@{}l@{}}2,475\\ 2,475\end{tabular}                      &     31,930     \\ \midrule
                        
        MED: Test A  & Culture &  \begin{tabular}[l]{@{}l@{}}IHI \\ Tanzania\end{tabular}   & \begin{tabular}[l]{@{}l@{}}Bednet \\ (2020)\end{tabular}      & \begin{tabular}[l]{@{}l@{}}Phone\\ 8\,kHz\end{tabular}           & \begin{tabular}[l]{@{}l@{}}4,118\\ 4,118\end{tabular}                   & 3,979   \\ 
                  \rowcolor{white}
         MED: Test B  & Culture                  & \begin{tabular}[l]{@{}l@{}} OxZoology\\ UK\end{tabular}     & \begin{tabular}[l]{@{}l@{}}Cage \\ (2016)\end{tabular}      & \begin{tabular}[l]{@{}l@{}}Phone\\ 8\,kHz\end{tabular}                                                              & \begin{tabular}[l]{@{}l@{}}737 \\ 737\end{tabular}                   &  2,307    \\ \midrule   \rowcolor{white}
                            
\multicolumn{5}{g}{\textbf{Total}}                                                      & \begin{tabular}[l]{@{}l@{}}\textbf{71,286}\\ \textbf{64,843}\end{tabular}                      &    \textbf{53,227}      \\ \bottomrule
\end{tabular}
\end{table}
\FloatBarrier

\paragraph{Are there any errors, sources of noise, or redundancies in the
dataset?} To our knowledge there are no redundancies, duplicate files, corrupt files or unintended bugs. Despite comprehensive manual checks, label errors due to human entry and ambiguity in sound type may remain.

\paragraph{Is the dataset self-contained, or does it link to or otherwise rely on
external resources (e.g., websites, tweets, other datasets)?}
The data is self-contained, generated from a PostgreSQL database which is hosted on University of Oxford servers. The data itself is hosted on Zenodo, and the code is accessible on GitHub.
\paragraph{Does the dataset contain data that might be considered confidential?} No, explicit permission was obtained where speech is present.
\paragraph{Does the dataset contain data that, if viewed directly, might be offensive, insulting, threatening, or might otherwise cause anxiety?} The audio recordings of mosquitoes may cause distress or discomfort to individuals with medical issues that pertain to mosquito sound.
\paragraph{Does the dataset identify any subpopulations (e.g., by age, gender)?}
The metadata identifies subpopulations of species complexes by species, and further by gender, age and plurality type (for example, if there was more than one mosquito recorded at a label). Further discriminating factors are described in Appendix \ref{sec:appendix_db_metadata}.
\paragraph{Is it possible to identify individuals (i.e., one or more natural persons), either directly or indirectly (i.e., in combination with other
data) from the dataset?} Yes, the speakers may announce the recording ID at the start of a recording, however explicit consent was obtained. It may be possible to trace to the person conducting the experiment indirectly.
\subsection{Collection Process}
\label{sec:appendix_datasheet_collection}
\paragraph{How was the data associated with each instance acquired?}
The data was collected globally at numerous research facilities. We summarise the data collection efforts below:

\begin{itemize}
\item \textbf{UK, Kenya, USA}:
Recordings from laboratory cultures at the London School of Tropical Medicine and Hygiene (LSTMH), the United States Army Medical Research Unit-Kenya (USAMRU-K); Center for Diseases Control and Prevention (CDC), Atlanta as well as with mosquitoes raised from eggs at the Department of Zoology, University of Oxford. Mosquitoes were recorded by placing a recording device into the culture cages where one or multiple mosquitoes were flying, or by placing individual mosquitoes into large sample cups and holding these close to the recording devices.

\item \textbf{Tanzania i)}:
Mosquitoes recorded at Ifakara Health Institute's semi-field facility (\textit{`Mosquito City'}) at Kining'ina. The facility houses six chambers containing purpose-built experimental huts, built using traditional methods and representing local housing constructions, with grass roofs, open eaves and brick walls. Four different configurations of the HumBug Net, each with a volunteer sleeping under the net, were set up in four chambers. Budget smartphones were placed in each of the four corners of the HumBug Net. Each night of the study, 200 laboratory cultured \textit{An.} \textit{arabiensis} were released into each of the four huts and the MozzWear app began recording.
\item \textbf{Tanzania ii)} A collection and recording project in the Kilombero Valley, Tanzania. HBNs, larval collections and CDC-LTs were used to sample wild mosquitoes and record them in sample cups in the laboratory. \textit{Anopheles} \textit{gambiae} and \textit{An. funestus} (another highly dangerous mosquito found across sub-Saharan Africa), are also siblings within their respective species complexes. Thus, standard PCR identification techniques \citep{scott1993identification} were used  to fully identify mosquitoes from these groups. The Tanzanian sampling has collected 17 different species including: \textit{An. arabiensis} (a member of the \textit{gambiae} complex), \textit{An. coluzzii}, \textit{An. funestus}, \textit{An. pharoensis} (see Appendix \ref{sec:appendix_db_metadata}, Figure \ref{fig:species_dist} for a full breakdown).

\item \textbf{Thailand}:
Mosquitoes were sampled using ABNs, HBNs and larval collections over a period of two months during peak mosquito season (May to October 2018). Sampling was conducted in Pu Teuy Village (Sai Yok District, Kanchanaburi Province, Thailand) at a vector monitoring station owned by the Kasetsart University, Bangkok. The mosquito fauna at this site include a number of dominant vector species, including \textit{An. dirus} and \textit{An. minimus} alongside their siblings (\textit{An. baimaii} and \textit{An. harrisoni}) respectively (Appendix \ref{sec:appendix_db_metadata}, Figure \ref{fig:species_dist} gives a species histogram for this dataset).
Sampling ran from 6 pm to 6 am, as most anopheline vectors prefer to bite during the night. Mosquitoes were collected at night, carefully placed into large sample cups and recorded the following day using the high-spec Telinga field microphone and a budget smartphone.
\end{itemize}

\paragraph{What mechanisms or procedures were used to collect the data
(e.g., hardware apparatus or sensor, manual human curation, software program, software API)?} A summary of equipment is as follows:
\begin{itemize}
    \item Smartphone (Itel, Alcatel, and others) audio recording with the MozzWear application. Smartphone devices may have variable sample rates, as denoted by the sample rate column of the metadata. The version of MozzWear used in the curation of this dataset recorded audio in 8,000\,Hz mono wave format.
    \item Telinga EM-23 field microphone, and Tascam, Olympus recording devices recording at 44,100\,Hz. The Telinga is a very sensitive, low-noise microphone which was widely adopted in bioacoustic studies.
    \item Human labelling with Excel.
    \item Human labelling with Audacity (GNU GPLv2 license).
    \item Labels produced by a Bayesian convolutional neural network (our own, MIT license, included in paper).
    \item Voice activity detection and removal with WebRTC (BSD license).
    \item Python (BSD-style license), MongoDB (Server Side Public License), Django (BSD license), Apache (GPLv3 license), PostgreSQL (BSD/MIT-like license), Unix for databases, HTML dashboards, and post-processing.
\end{itemize}
\paragraph{Who was involved in the data collection process (e.g., students,
crowdworkers, contractors) and how were they compensated (e.g.,
how much were crowdworkers paid)?} Researchers from the locations previously mentioned, paid salary from their respective institutions,  through the grants disclosed previously.
\paragraph{Over what timeframe was the data collected?}
2015 to 2020 (and ongoing).
\paragraph{Were any ethical review processes conducted (e.g., by an institutional review board)?}
We have obtained the ethics approval from the following committees:
\begin{itemize}
\item Oxford Tropical Research Ethics Committee (OxTREC Ref. 548-19) -- University of Oxford (UK).
\item       Ifakara Health Institute (IHI)-IRB -- Tanzania
\item          National Institute for Medical Research -- Tanzania
\item          School of Public Health at the University of Kinshasa (KSPH) -- DRC
\end{itemize}
\subsection{Preprocessing/cleaning/labeling}
\label{sec:appendix_datasheet_preprocessing}
\paragraph{Was any preprocessing/cleaning/labeling of the data done (e.g.,
discretization or bucketing, tokenization, part-of-speech tagging,
SIFT feature extraction, removal of instances, processing of missing values)?}
The data underwent rigorous curation, from manual adjustment to labels supplied in text files, to commands in the database to deal with incorrectly entered label times resulting in missing data. To encourage reproducibility and compatibility for future data release, all the label and audio quality control is performed before uploading to the database, and within the dataset itself.

Example of quality control code to check that the label end does not exceed the length (which happens frequently when labels are entered by hand into Audacity with end times longer than the recording and then exported to a text file):

\begin{lstlisting}[
           language=SQL,
           showspaces=false,
           breaklines = true,
           numbers=left,
           numberstyle=\tiny,
           morekeywords={\copy},
           commentstyle=\color{gray}
        ]
SELECT path, fine_start_time, fine_end_time, sound_type, length
FROM label
LEFT JOIN mosquito ON (label.mosquito_id = mosquito.id)
RIGHT JOIN audio ON (label.audio_id = audio.id)
RIGHT JOIN location ON (audio.loc_id = location.id)
WHERE fine_end_time > length;
\end{lstlisting}

Sources with low estimated label quality were either removed or manually re-labelled and amended in the database.

\paragraph{Was the “raw” data saved in addition to the preprocessed/cleaned/labeled
data (e.g., to support unanticipated future uses)?}
Yes, all data that may have future utility (and has not been yet used for that purpose) has been released. Unprocessed, and currently unlabelled data is also all stored on the database server, but requires further curation and data entry to the specific data tables before release. We plan to periodically update the database as more data becomes available.
\paragraph{Is the software used to preprocess/clean/label the instances available?}
The software to do so included Audacity, PostgreSQL, Python, Excel, and is available and well-maintained. We will make use of it in future for future data curation.

\subsection{Uses}
\label{sec:appendix_datasheet_uses}
\paragraph{Has the dataset been used for any tasks already?}
A subset of this data (recorded in Thailand, Kenya, UK, USA) has been used to train a machine learning model to distinguish and detect a mosquito from its acoustic signature. The model was a 4-layer Bayesian convolutional neural network implemented in Keras. The predictive entropy and mutual information were used to screen predictions over thousands of hours of data. Hand labels were added to correct predictions, and the labels were fed back into the database \citep{2021pkddBNN}. Code for the training and resulting predictive pipeline is available on \url{https://github.com/HumBug-Mosquito/MozzBNN}. 

Other past use cases and publications can be found in related works from the link of the following section. We summarise these here as:
\begin{itemize}
    \item Bioacoustic classification with wavelet-conditioned neural networks \citep{kiskin2017mosquito,kiskin2018bioacoustic}.
    \item Cost-sensitive mosquito detection  \citep{li2017cost}
    \item A case study of species classification with field recordings \citep{lidcasefast}
    \item A release of a subset of this database for crowdsourcing (with baseline mosquito detector model) \citep{kiskin2019data}
\end{itemize}
\paragraph{Is there a repository that links to any or all papers or systems that use the dataset?}
 Yes, the Zenodo data directory \url{https://zenodo.org/record/4904800} contains all the references to projects, papers and code which are associated with this dataset.
\paragraph{ What (other) tasks could the dataset be used for?}
 A list of use cases is not limited to, but may include:
\begin{enumerate} 
\item \textbf{Validating species classification models from the literature.} 
\item \textbf{Frequency analysis.} Identifying the fundamental and harmonic frequencies of flight tone for a particular species, to improve upon the understanding of bioacoustics literature, and entomological research.
\item \textbf{Examining inter-species (or similar) variability.} For example, the effect on the sound of flight as a result of age, gender, or any field supported in the database.
\end{enumerate}

We now expand upon each point:

\begin{enumerate}

    \item \textbf{Validating species classification models from the literature.} As a result of procuring curated data with species meta-information of both wild and lab mosquitoes, this dataset serves as an ideal test-bed to verify the effectiveness of existing species classification approaches. We encourage researchers to validate their models by making use of these data to form their own test sets without re-training their models on any parts of this dataset. Strong species discrimination performance would signify a great opportunity to utilise acoustics as a wide-scale surveillance tool. 
    
    It would also be very useful to examine transfer learning approaches, where previous models are re-trained and tested on the suggested splits of the data for either task.
    If you encounter any issues, or require further information do not hesitate to contact the database maintainers (Appendix \ref{sec:appendix_database_maintenance}).
    
    \item \textbf{Frequency analysis.} Earlier works in the literature proposed more hand-crafted approaches to building detection or classification models. These may be especially useful in very low-power embedded devices which require fast and efficient algorithms. These approaches were typically centered around specific harmonic inter-peak ratios (See \citet[Sec.\,3.2]{kiskin2020machine} for an overview of relevant prior work). Frequency analysis may be performed on any parts of this dataset, including on species which are under-represented. In particular, the CDC dataset contains a wide range of unique species which are sparsely labelled, however the labelled sections have very high signal-to-noise ratio. As with previous suggested use cases, we recommend trialling approaches on disjoint sets of experiments (or at the very least individual mosquito recording within an experimental set). Once again, there exists an excellent opportunity to validate models from the literature on their ability to distinguish species on this dataset.
    
    \item \textbf{Examining the effect of species variability on their flight tone.} It is well known that mosquitoes exhibit significant variability in their physical (and therefore acoustic) properties within a species. These occur due to a multitude of factors including the age, wingspan, gender. Additionally, confounding factors such as the temperature, humidity, and potentially their fed status, can increase the difficulty in distinguishing individuals within and across species. As we maintain as much metadata as possible, this dataset provides the opportunity to examine such factors. In future releases, temperature and humidity will be added where possible, and this data is expected to be available in an update on the Tanzanian cup recordings which has already good metadata coverage including \texttt{species}, \texttt{age}, \texttt{gender}, \texttt{fed}, \texttt{method}. If you wish to have early access to additional metadata, please contact us and we will make the availability of such metadata a higher priority.
    
\end{enumerate}

\paragraph{Is there anything about the composition of the dataset or the way it was collected and preprocessed/cleaned/labeled that might impact future uses?}
No, the dataset is specifically organised in PostgreSQL in a way to be consistent with future release. However, in future more metadata may become available for legacy datasets, and larger subsets may become available upon addition of labels.

\paragraph{Are there tasks for which the dataset should not be used?}
No.

\subsection{Data bias}
\label{sec:appendix_datasheet_databias}
Data is collected with varying recording paradigms, and is sampling a broad (and not fully understood) population of mosquitoes. This induces inherent biases which may affect an algorithm's performance when acting as either a mosquito detector or species discriminator. We attempted to capture  biases as well as possible with comprehensive metadata coverage, which we encourage users to explore for their own use cases. We discuss potential sources of bias and suggest mitigation strategies as follows:

\begin{enumerate} 
\item \textbf{Nature of mosquitoes: lab or wild}. These are denoted by \texttt{location\_type}. The controlled conditions of laboratory cultures produce uniformly sized fully-developed adult mosquitoes which are used for a variety of purposes, including trialling new insecticides or examining the genome of these insects.  Models trained on purely lab cultures run the risk of overfitting to this artificial subpopulation, encountering difficulty when generalising even to the same species. Wild mosquitoes on the other hand exhibit greater variation, at the cost of a much more laborious collection procedure. When constructing models, it is advised to train on wild data, but caution needs to be taken when testing on mosquitoes from uniform subpopulations. 

\item \textbf{Location bias.} Distributions of mosquitoes may vary across regions, which would be an interesting avenue for further work. This database is expected to form training data for an upcoming challenge, where a hidden test set will consist of recordings conducted at the DRC, carried out with as similar recording methodology to Tanzanian cup data used in MSC. At that stage it will be possible to consider in more depth whether the location is a factor causing significant bias and therefore difficulty for the models.

\item \textbf{Recording device} corresponding to metadata from \texttt{device}, It is crucial that datasets are not constructed in a way where one device is used for only positive or negative instances (e.g. all noise is from one device, and all mosquito from another). If trained in such a manner, it will be easy for a high-dimensional model such as a neural network to learn the characteristics of the microphone response and use this confounding factor for classification. To mitigate this, we have included a negative control group for each experiment, and therefore also device. This issue becomes especially critical for species classification, where different species may be captured with different devices. Careful consideration and construction of data with the use of the device metadata will help avoid, or at the very least alert to possible confounding. If it is not possible to control the device, it may be desirable to use features which are (more) invariant to microphone type, e.g. MFCCs or high-level pre-trained feature representations such as VGGish embeddings \citep{gemmeke2017audio}.
\item \textbf{Data imbalance.} Biased models for either species classification or mosquito detection may arise when trained naively without balancing distributions of species, or positive and negative samples. In the case of mosquito detection, a predominance of one species will likely increase the model's ability to detect mosquitoes of that particular species, while performing worse on less well-represented groups. This is a potential source of improvement worth investigating, especially for the data split suggested in Table \ref{tab:Summary_for_appendix}. Additionally, a closer look at species-specific performance may reveal areas for further model improvement. We recommend benchmarking against the baselines supplied to investigate areas of improvement.

For our MSC tasks we weight class samples by the inverse of their frequency. There are however well-known drawbacks to this. Bayesian models which take into account asymmetrical cost functions aim to alleviate this problem \citep{cobb2018loss}. A further option is to use different step functions in the data partitioning/augmentation pipelines. A starting point would be to modify \texttt{step\_size} in \texttt{feat\_util.py} in  a class-specific function, to artificially balance the relative frequency of data samples of desired classes. We have found oversampling or undersampling to produce worse per-class ROC than inverse weighting, however there is room for improvement in future work.
\end{enumerate}

\subsection{Database maintenance}
\label{sec:appendix_database_maintenance}
\paragraph{Who is supporting/hosting/maintaining the dataset?}
Please contact Dr. Ivan Kiskin at \texttt{ivankiskin1@gmail.com}, who is maintaining the dataset. Alternative contacts include Professor Steve Roberts at \texttt{sjrob@robots.ox.ac.uk} at the University of Oxford Machine Learning Research Group.

\paragraph{How will the dataset be updated and what is the maintenance policy?}
The data will be updated as new data and/or metadata from new trials is obtained and curated. We expect the following updates:

\begin{enumerate}
    \item Recordings of wild captured mosquitoes in DRC:
\begin{itemize}
    \item \textbf{Date:} Q2/Q3 2021
    \item \textbf{Summary:} 15 species, at minimum 2000 wild captured individual mosquitoes
    \item \textbf{Metadata:} species (with PCR identification where appropriate), gender, fed status, temperature, humidity, collection method, recording device information, time of collection
\end{itemize}
\item Additional metadata for IHI Tanzanian cup recordings:
\begin{itemize}
    \item \textbf{Date:} Q2/Q3 2021
    \item \textbf{Summary:} Additional metadata
    \item \textbf{Metadata:} Temperature, humidity, wing span images (and wing lengths)
\end{itemize}
\end{enumerate}

To ensure the documentation is up to date, any additional metadata and code will be documented in this supplement and the datasheet for datasets. The supplement is available on GitHub alongside the baseline code at \url{https://github.com/HumBug-Mosquito/HumBugDB/tree/master/docs}. Database revisions will be incremented in the format of X.X.X (current version 0.0.1). The main url,  \url{https://zenodo.org/record/4904800}, will always resolve to the latest version. If you intend to use a specific version you may select the version from the main page. Both the data and metadata are fully supported with versioning.

Updates will be communicated as follows:
\begin{itemize}
    \item GitHub commits (mailing list), and releases.
    \item Posts on the HumBug official twitter account \url{https://twitter.com/oxhumbug}.
    \item Updates on our official website on \url{https://humbug.ox.ac.uk/news/}.
    \item Follow-up publications utilising additional data (arXiv and proceedings where appropriate).
\end{itemize}

\paragraph{If others want to extend/augment/build on/contribute to the
dataset, is there a mechanism for them to do so? If so, please
provide a description.}
If you would like to contribute to this data, please contact the database host and supervising professor. We would be happy to curate data and provide requirements which would help qualify a dataset for hosting. All contributions will be credited appropriately in future work.